\documentclass[linenumbers]{aastex631}

\usepackage{CJK}
\usepackage{color}
\usepackage{soul}
\usepackage{natbib}
\usepackage{amsmath}
\usepackage{xspace}
\usepackage{lineno}
\linenumbers
\usepackage{graphicx}
\usepackage{enumitem}
\usepackage{amssymb}
\usepackage{xifthen}
\usepackage{todonotes}
\usepackage[normalem]{ulem}
\usepackage{pifont}%

\hypersetup{    
  colorlinks      = {true},
  linkcolor       = {blue},
  citecolor       = {blue},
  urlcolor        = {blue},
}

\newcommand{\code}[1]{\texttt{#1}\xspace}

\newcommand{\unit}[1]{\ensuremath{\mathrm{\,#1}}\xspace}

\newcommand{\teff}{\ensuremath{T_\mathrm{eff}}\xspace}
\newcommand{\logg}{\ensuremath{\log\,g}\xspace}
\newcommand{\alphafe}{\unit{[\alpha/Fe]}}

\newcommand{\Gaia}{\emph{Gaia}\xspace}

\newcommand{\kms}{\unit{km\,s^{-1}}}

\usepackage{comment}

\shorttitle{Chemical Abundances of Nyx}
\shortauthors{Wang et al.}

\graphicspath{{./}{figures/}}

\begin{document}
\nolinenumbers
\begin{CJK*}{UTF8}{gbsn}

\title{High-Resolution Chemical Abundances of the Nyx Stream}

\correspondingauthor{Shuyu Wang}
\email{shuyuwang.astro@gmail.com}

\author[0000-0001-6663-5605]{Shuyu Wang (汪书玉)}
\affiliation{Department of Astronomy \& Astrophysics, University of Chicago, 5640 S Ellis Avenue, Chicago, IL 60637, USA}
\affiliation{Kavli Institute for Cosmological Physics, University of Chicago, Chicago, IL 60637, USA}

\author[0000-0003-2806-1414]{Lina Necib}
\affiliation{Department of Physics and Kavli Institute for Astrophysics and Space Research,
Massachusetts Institute of Technology,
77 Massachusetts Ave, Cambridge MA 02139, USA}
\affiliation{The NSF AI Institute for Artificial Intelligence and Fundamental Interactions, 77 Massachusetts Ave, Cambridge MA 02139, USA}

\author[0000-0002-4863-8842]{Alexander~P.~Ji}
\affiliation{Department of Astronomy \& Astrophysics, University of Chicago, 5640 S Ellis Avenue, Chicago, IL 60637, USA}
\affiliation{Kavli Institute for Cosmological Physics, University of Chicago, Chicago, IL 60637, USA}

\author[0000-0002-4669-9967]{Xiaowei Ou}
\affiliation{Kavli Institute for Astrophysics and Space Research,
Massachusetts Institute of Technology,
77 Massachusetts Ave, Cambridge MA 02139, USA}

\author[0000-0002-8495-8659]{Mariangela Lisanti}
\affiliation{Department of Physics, Princeton University, Princeton, NJ 08544, USA}
\affiliation{Center for Computational Astrophysics, Flatiron Institute, 162 Fifth Ave, New York, NY 10010, USA}

\author[0000-0002-4739-046X]{Mithi~A.~C.~de los Reyes}
\affiliation{Department of Physics, Stanford University, 382 Via Pueblo Mall, Stanford, CA 94305, USA}
\affiliation{Kavli Institute for Particle Astrophysics \& Cosmology, P.O. Box 2450, Stanford University, Stanford, CA 94305, USA}

\author[0000-0001-6369-1636]{Allison~L.~Strom}
\affiliation{Department of Physics and Astronomy and Center for Interdisciplinary Exploration and Research in Astrophysics (CIERA), Northwestern University, 2145 Sheridan Road, Evanston, IL 60208, USA}

\author[0000-0001-9812-4957]{Mimi Truong}
\affiliation{California State University, Northridge
18111 Nordhoff St, Northridge, CA 91330}

\begin{abstract}
\nolinenumbers
Nyx is a nearby, prograde, and high-eccentricity stellar stream physically contained in the thick disk but with an unknown origin.
Nyx could be the remnant of a disrupted dwarf galaxy, in which case the associated dark matter substructure could affect terrestrial dark matter direct detection experiments. Alternatively, Nyx could be a signature of the Milky Way's disk formation and evolution.
To determine the origin of Nyx, we obtained high-resolution spectroscopy of 34 Nyx stars using Keck/HIRES and Magellan/MIKE.
A differential chemical abundance analysis shows that most Nyx stars reside in a metal-rich ($\mbox{[Fe/H]} > -1$) high-$\alpha$ component that is chemically indistinguishable from the thick disk. This rules out an originally suggested scenario that Nyx is the remnant of a single massive dwarf galaxy merger.
However, we also identify five substantially more metal-poor stars ($\mbox{[Fe/H]} \sim -2.0$) that have chemical abundances similar to the metal-weak thick disk.
It remains unclear how stars chemically identical to the thick disk can be on such prograde, high-eccentricity orbits. 
We suggest two most likely scenarios: that Nyx is the result of an early minor dwarf galaxy merger or that it is a record of the early spin-up of the Milky Way disk---although neither perfectly reproduces the chemodynamic observations.
The most likely formation scenarios suggest that future spectroscopic surveys should find Nyx-like structures outside of the Solar Neighborhood.

\end{abstract}

\keywords{Stellar abundances; Stellar kinematics; Milky Way formation; Dark matter}

\section{Introduction} 
\label{sec:intro}

The standard $\Lambda$ Cold Dark Matter paradigm predicts that galaxies form hierarchically \citep[e.g.,][]{White&Rees_1978}, such that massive galaxies like the Milky Way grow through the accretion of smaller galaxies.
As the Milky Way accretes smaller satellite galaxies, its gravitational potential tidally disrupts these satellites, which can leave behind streams of stars \citep[e.g.,][]{1998ApJ...495..297J} as well as dark matter \citep[e.g.,][]{Read_2008,2018PhRvL.120d1102H,2018JCAP...04..052H} in the Galaxy.
The stellar component of the streams is crucial in understanding the formation of the Milky Way. 
The dark matter component, or the dark matter substructure, could affect the dark matter phase space distribution and thus could have profound implications for the terrestrial direct detection of dark matter.
Stream stars retain key information of the chemical abundances and kinematics of their progenitor galaxies long after the accretion events~\citep[e.g.,][]{1997ARA&A..35..503M,Helmi_1999,Freeman2002,Venn_2004,Bullock_2005,Robertson2005,Font_2006,2008A&ARv..15..145H,Helmi_2020}. Hence, detailed chemodynamic analyses of stars in the tidal debris can be used to identify the origin of a stellar stream, to reconstruct the merger event with the Milky Way, and possibly, to determine the presence of a dark matter substructure accompanying such a merger within the Galaxy.

The advent of all-sky measurements of stellar proper motions from \Gaia \citep{2016A&A...595A...1G,2016A&A...595A...2G,Gaia_DR2,2021A&A...649A...1G} has resulted in the discovery of numerous new kinematic structures in the Milky Way \citep[e.g.,][]{Helmi_2020,Naidu_2020,Yuan_2020}. These Galactic building blocks include the debris flow referred to as the Gaia-Sausage Enceladus~(GSE) \citep{Belokurov_2018, Helmi_2018, Mackereth_2019, Haywood_2018}, as well as streams such as the Helmi stream \citep{Helmi_1999}, Sequoia \citep{Myeong_2019}, and Thamnos \citep{Koppelman_2019}. 

One of the most recent discoveries is Nyx, a prograde stellar stream in the Solar vicinity~\citep{2020NatAs...4.1078N}. Nyx was found after applying clustering algorithms to a catalog of likely accreted stars from the second data release~(DR2) of \emph{Gaia}~\citep{2020A&A...636A..75O}. To build this catalog, \citet{2020A&A...636A..75O} trained a neural network on both simulation and data, using stellar 5D kinematics to distinguish between an accreted or \emph{in-situ} origin. First, the network was trained on the synthetic \Gaia catalog \texttt{Ananke}~\citep{2020ApJS..246....6S},
based on Milky Way-like galaxies from the \texttt{Latte} suite of simulations~\citep{2016ApJ...827L..23W}, which uses the Feedback In Realistic Environments (FIRE) code for the hydrodynamic implementation~\citep{2014MNRAS.445..581H,2015MNRAS.450...53H,Hopkins_2018}. 
Then, transfer learning~\citep{caruana1994learning,bengio2012deep} was performed using the RAVE~DR5-\Gaia~DR2 cross-match~\citep{Kunder_2017}. At the end of the training, each \Gaia DR2 star received a score $S \in [0,1]$, where $1$~($0$) means the star is likely accreted~(\emph{in-situ}). In the resulting accreted catalog, \citet{2020ApJ...903...25N} rediscovered several known stellar substructures near the Sun, in addition to Nyx. The latter stands out as an overdensity of stars confined within 3~kpc of the disk plane that are on prograde orbits, but with large radial velocities (i.e., high eccentricities). The Nyx stream has since been identified with alternate methods
\citep[e.g.,][]{Donlon_2019, Donlon2021, Gryncewicz_2021}. 

There are two classes of explanations for the origin of the Nyx stream with disparate effects on the terrestrial direct detection experiments of dark matter. These two classes revolve around Nyx having an accompanying dark matter component, which reduces to whether or not Nyx is the remnant of one or more dwarf galaxy accretion events
\citep{Abadi_2003,Sales_2009,Read_2009,Read_2008,Purcell_2009,Ling_2010,Pillepich_2014,Rodriguez_Gomez_2017}.
If Nyx is indeed a remnant of a merger, it would potentially contribute to a dark matter substructure which could affect the local dark matter phase-space distribution and impact constraints from direct detection experiments \citep{Read_2009,Bruch09}.
Alternatively, Nyx could be a kinematic perturbation to the thick disk \citep[e.g.,][]{van_Donkelaar_2022,Roskar_2008}. This formation scenario would not contribute to the local dark matter phase-space distribution. 

Chemical abundances of Nyx stars can distinguish between these formation scenarios, since chemical evolution causes stars from dwarf galaxies to have distinct chemistry from the Milky Way \citep[e.g.,][]{Venn_2004}. In particular, if Nyx is the remnant of an accreted dwarf galaxy, it should have chemical composition similar to dwarf galaxies, e.g. low $\alpha$ abundances and evidence of the metal-poor s-process.
By cross-matching with the RAVE-on chemical abundance catalog \citep{Kunder_2017,Casey_2017}, \citet{2020NatAs...4.1078N} found three metal-rich Nyx stars with Mg abundances $\mbox{[Mg/Fe]} \lesssim 0.3$, which may be consistent with both the thick disk and accreted dwarf galaxy formation scenarios. The inconclusive chemical abundance results arising from a combination of small sample size and large abundance errors motivate higher precision chemical abundance analysis of a larger sample of Nyx stars.

Recently, \cite{Zucker_2021} analyzed the chemical abundances of 18 Nyx candidates from GALAH~DR3 \citep{Buder_2021} and 9 candidates from APOGEE~DR16 \citep{APOGEE_DR16}. The GALAH~DR3 sample had $\mbox{[Fe/H]} > -0.9$. Though the APOGEE~DR16 sample mostly overlapped in this range, \cite{Zucker_2021} identified two low-metallicity stars with $\mbox{[Fe/H]} \sim -1.5$. They compared both samples with kinematically-selected thick disk stars, concluding that Nyx is chemically consistent with the thick disk. \cite{Horta_2022} examined chemical abundances of 589 stars consistent with Nyx kinematics from the APOGEE~DR17 survey~\citep{APOGEE_DR17}. They identified a low-metallicity tail with $\mbox{[Fe/H]} < -1.5$ in the sample, but did not provide further explanations for its existence. They also considered Nyx to be chemically similar to the \emph{in situ} high-$\alpha$ thick disk. Both of these recent results suggest that Nyx could be a kinematic perturbation to the thick disk, though neither provides a plausible explanation for the existence of the metal-poor tail. This motivates further chemodynamic studies of the low-metallicity tail to unveil the origin of the Nyx stream.

In this paper, we study the chemodynamics of 34 Nyx stars observed with the Keck/HIRES and Magellan/MIKE spectrographs. We perform the first detailed analysis of the chemical abundances of the Nyx stars using a differential abundance analysis. Overall, we find that most Nyx stars have similar chemical abundances to the thick disk, consistent with previous studies. However, we also identify five stars that populate a highly-eccentric metal-poor tail ($\mbox{[Fe/H]}\sim-2.0$) which is not typically associated with the thick disk. This low-metallicity tail could be key to unveiling the origin of the Nyx stream, be it the remnant of multiple dwarf galaxy mergers or an unusual kinematic structure formed in the early turbulent Galactic disk.

Throughout this paper, we use $v_r$, $v_{\phi}$ and $v_{\theta}$ to denote Galactocentric velocities in spherical coordinates. The paper is organized as follows: Sec.~\ref{sec:obs} provides details of the observation and target selection. Sec.~\ref{sec:abund} introduces the methods used to analyze the chemical abundances of Nyx stars. Sec.~\ref{sec:result} presents the chemical abundance results, comparing to a thick disk sample from \citet{Bensby_2014} and highlighting the fact that Nyx contains a metal-poor tail. Sec.~\ref{sec:dynamics} studies the kinematics of the Nyx stars. Sec.~\ref{sec:discussion} discusses possible origins of the Nyx stream as well as their implications on the local dark matter phase-space distribution. Sec.~\ref{sec:conclusion} concludes the paper.

\end{CJK*}

\section{Target Selection and Observations}
\label{sec:obs}

To identify accreted substructures, \citet{2020ApJ...903...25N} performed a Gaussian mixture model analysis on stellar velocities in a catalog of accreted stars with 
neural network score $S > 0.95$ from \citet{2020A&A...636A..75O} (see Sec.~\ref{sec:intro} for a summary). 
The Nyx stars are characterized by a Gaussian velocity distribution with mean
$\{v_r, v_{\phi}, v_{\theta}\} = \{134, 130, 53.0\} \kms$ and dispersion $\{\sigma_r, \sigma_{\phi}, \sigma_{\theta}\} = \{67.2, 45.8, 66.3\} \kms$ \citep{2020NatAs...4.1078N}. 
We select 34 out of 94 stars from this velocity cluster with the highest probabilities of belonging to the Nyx velocity cluster. 
This biases our sample towards larger velocities and a smaller velocity dispersion compared to the full Nyx sample. The mean velocities of the Nyx sample studied in this paper are $\{v_r, v_{\phi}, v_{\theta}\} = \{167, 172, 44.4\} \kms$, with dispersions $\{\sigma_r, \sigma_{\phi}, \sigma_{\theta}\} = \{55.0, 30.9, 59.0\} \kms$.

\begin{figure}
\begin{center}
\includegraphics[width=0.85\textwidth]{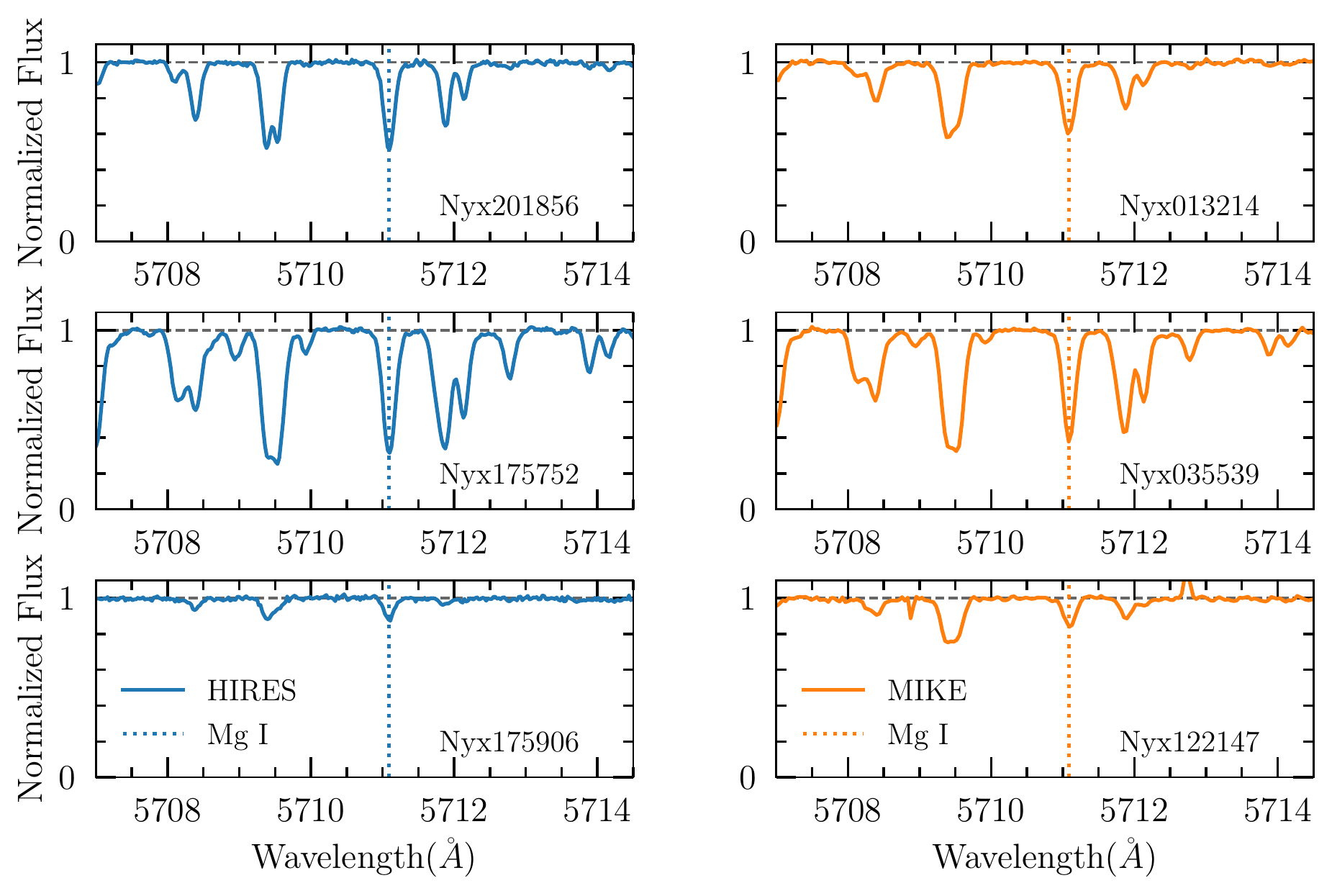} \\
\end{center}
\caption{Spectra of selected Nyx stars observed with HIRES (blue) or MIKE (orange) around the Mg~I line at 5711~\AA (dashed vertical lines). Stars are selected and sorted based on star type and metallicity. From top to bottom, the star in each subplot is a metal-rich dwarf, a metal-rich giant or a metal-poor star.
}
\label{fig:spectra}
\end{figure}

We obtained high-resolution spectroscopy of these 34 Nyx stars: 28 stars on August 1, 2020 with the $0\farcs5$ slit using the Keck/HIRES spectrograph ($R\sim \mathrm{67k}$ at 4100~\AA, \citealt{Vogt_1994}) and 6 stars on January 3, 2021 or July 28, 2021 (Nyx122147) with the $0\farcs5$ slit using the Magellan/MIKE spectrograph ($R\sim \mathrm{50k/40k}$ on the blue/red arm of MIKE; \citealt{Bernstein_2003}). 
The HIRES spectra range from 3650 to 8160~\AA, and the MIKE spectra range from 3300 to 9400~\AA. The portion of the spectra used for abundance analysis ranges from 4000 to 7000~\AA. Figure \ref{fig:spectra} shows spectra around the Mg~I line at 5711~\AA~for a few metal-rich dwarf, metal-rich giant and metal-poor Nyx stars observed with HIRES and MIKE.
Data for the 28 Nyx stars observed with HIRES were reduced with MAKEE~v6.4\footnote{\url{https://sites.astro.caltech.edu/~tb/ipac_staff/tab/makee/}}, while data for the 6 stars observed with MIKE were reduced with CarPy \citep{Kelson_2003}.

We use the thick disk stars from \citet{Bensby_2014} as a comparison chemical and kinematic sample. \citet{Bensby_2014} obtained high-resolution spectroscopy of 714 F and G dwarf stars in the Solar Neighborhood that traced multiple nearby kinematic components of the Milky Way (thin disk, thick disk, stellar halo, and streams). The selection function for these stars was purposefully biased to probe as many Galactic components and the widest metallicity ranges of such components as possible. These stars were kinematically classified into thin disk, thick disk, and halo stars using the Gaussian velocity distributions defined in \citet{Bensby2003}.  
We select 201 thick disk stars
with thick disk to halo ratio $\rm{TD/H} > 2$ and thick disk to disk ratio $\rm{TD/D} > 2$.\footnote{Changing the cut on $\rm{TD}/\rm{H}$ from 2 to 10
only removes 6 stars out of 201 from the thick disk sample.} 
We expect there to be six contaminating stars (3.0\%) from the halo in this sample. 
Cross-matching these thick disk stars with the accreted catalog from \citet{2020A&A...636A..75O}, we find that the neural network scores of these thick disk stars is $S < 0.05$. 
Given the difference in scores, we expect there to be no contamination from kinematically ordinary thick disk stars in our 34-star Nyx sample.

We will determine chemical abundances differentially (see~Sec.~\ref{sec:diff}), so we select two stars from the \citet{Bensby_2014} thick disk sample to use as abundance references---the metal-rich reference star
HIP88622 and the metal-poor reference star HIP7162. HIP88622 was observed on August 1, 2020 with the $0\farcs5$ slit using the Keck/HIRES spectrograph (signal-to-noise ratio per pixel~(SNR) $\sim250$ at 6500~\AA). HIP7162 was observed on June 30, 2022 with the $0\farcs5$ slit using the MIKE/Magellan spectrograph (SNR $\sim114$ at 6500~\AA). 
For the reference star Arcturus, high-resolution spectra were obtained from \citet{Hinkle_2000}.

Table~\ref{tab:obs} lists the name, \Gaia~DR3 source ID, RA/Dec, Observation Date, Instrument, \Gaia G magnitude, exposure time ($t_{\text{exp}}$), SNR at 6500~\AA, and heliocentric radial velocity ($v_{\rm{hel}}$) of the observed Nyx stars. The last two columns of Table~\ref{tab:obs} provide a star's neural network score~($S$) from \citet{2020A&A...636A..75O} as well as its probability of belonging to the Nyx velocity distribution from \citet{2020NatAs...4.1078N}. Radial velocities were measured by cross-correlating with a MIKE spectrum of the metal-poor giant HD122563 around Mgb. For stars observed with multiple exposures, we report the heliocentric velocity measured in the middle of the observation. The formal velocity uncertainties are 0.1--$0.2 \kms$, but past experience shows that a typical systematic error on the velocity due to wavelength calibration and template mismatches is $1.0 \kms$ \citep[e.g.,][]{Ji_2020a}.

\section{Abundance Analysis}
\label{sec:abund}
We determine the stellar abundances using the 2017 version of the 1D Local Thermodynamic Equilibrium (LTE) radiative transfer code MOOG\footnote{\url{https://github.com/alexji/moog17scat}} with scattering \citep{Sneden_1973,Sobeck_2011}. We use \code{SMHR}\footnote{\url{https://github.com/andycasey/smhr/tree/py38-mpl313}} to normalize and stitch echelle orders, fit equivalent widths, interpolate ATLAS model atmospheres \citep{Castelli_2004}, run MOOG, and perform spectral syntheses \citep[as described in][]{Casey_2014,Ji_2020}. As described in the following subsections, stellar parameters are determined spectroscopically, and final abundances are determined differentially relative to the reference stars from \citet{Bensby_2014} and \citet{McWilliam_2013}.

\subsection{Line List and Line Measurements}
\label{sec:lines}
Our line list is primarily adopted from \citet{Jonsson_2017}, \citet{Jonsson_2019a}, and \citet{Jonsson_2019b}, supplemented by Fe lines from \cite{Bensby_2005} (see Table~\ref{tab:lines}), Na and Y lines from \citet{McWilliam_2013}, and Ti, Ni and Y lines from \citet{Roederer_2018}. We individually examine each line, rejecting those with heavy blending issues, saturation or non-detection. 
For the 28 Nyx stars observed with Keck/HIRES, some lines fall into chip gaps. These include some Fe lines and some Y lines.

We use a combination of equivalent widths and spectral synthesis to determine the stellar abundances. 
For equivalent widths, we use \code{SMHR} to fit a Gaussian or Voigt profile multiplied by a linear continuum to each absorption line. The equivalent width measurements are individually examined to flag problematic lines, remove heavily blended lines, and correct inaccuracies in local continuum fitting.
We use equivalent widths to determine the abundances of Na~I, Mg~I, Si~I, Ca~I, Ti~I, Fe~I/II, Ni~I, Zn~I, Y~II and Ba~II.
For lines with hyperfine splitting (e.g. La~II), we use synthetic spectra to determine the abundances. 
\code{SMHR} finds the best fit synthetic spectra using chi-squared minimization (see details in \citealt{Ji_2020}). We individually inspect all synthetic fits to ensure accuracy.

\begin{longrotatetable}
\begin{deluxetable*}{lccccccccccc}
\tablecolumns{12}
\tablecaption{\label{tab:obs}Observations}
\tablewidth{800pt}
\tabletypesize{\footnotesize}
\renewcommand{\arraystretch}{1.0}
\tablehead{
\colhead{Name} & \colhead{Source id} & 
\colhead{RA} & \colhead{Dec} & 
\colhead{Obs Date} & \colhead{Instrument} & 
\colhead{$G$} & \colhead{$t_{\rm{exp}}$} & 
\colhead{SNR} & \colhead{$v_{\rm{hel}}$} &
\colhead{Score} & \colhead{Velocity}
\\ 
\colhead{} & \colhead{(\Gaia~DR3)} & 
\colhead{(degrees)} & \colhead{(degrees)} & 
\colhead{} & \colhead{} & 
\colhead{(mag)} & \colhead{(seconds)} &
\colhead{6500\AA} & \colhead{($\kms$)} &
\colhead{}& \colhead{Probability}
}
\startdata
NYX004259 & 367439335551722240  & 10.74511785  & 37.38043004  & 2020-08-01 & HIRES      & 10.74 & 350              & 151 & 24.6   & 0.97   & 0.77         \\
NYX010249 & 314182978031619328  & 15.70439781  & 32.12158685  & 2020-08-01 & HIRES      & 11.69 & 900              & 151 & 26.2   & 0.98   & 0.76         \\
NYX010842 & 306730075901451520  & 17.17457698  & 26.70320141  & 2020-08-01 & HIRES      & 11.70 & 900              & 159 & 27.1   & 0.96   & 0.78         \\
NYX013214 & 4916946665759118720 & 23.06001744  & $-51.90289781$ & 2021-01-03 & MIKE       & 10.18 & 1200             & 140 & $-16.2$  & 0.97   & 0.97         \\
NYX014107 & 317636063018105088  & 25.27974372  & 33.92092371  & 2020-08-01 & HIRES      & 11.21 & 550              & 144 & 27.1   & 0.97   & 0.61         \\
NYX014725 & 305238279140992896  & 26.85528438  & 32.42734287  & 2020-08-01 & HIRES      & 8.88  & 70               & 145 & 27.4   & 0.96   & 0.89         \\
NYX022545 & 5174773415097847296 & 36.43952571  & $-10.98901054$ & 2020-08-01 & HIRES      & 11.62 & 900              & 148 & 26.3   & 0.96   & 0.94         \\
NYX035539 & 4665853149532721664 & 58.91244399  & $-71.36549004$ & 2021-01-03 & MIKE       & 11.04 & 4085             & 232 & $-4.0$     & 0.95   & 0.95         \\
NYX044831 & 4665093769251640704 & 72.13076827  & $-62.48367570$ & 2021-01-03 & MIKE       & 8.91  & 400              & 169 & $-4.4$   & 0.98   & 0.96         \\
NYX063721 & 5497838651784528000 & 99.33851551  & $-54.82999871$ & 2021-01-04 & MIKE       & 10.52 & 1600             & 213 & 1.4    & 0.97   & 0.80         \\
NYX064943 & 5497618547596953472 & 102.43003976 & $-53.84094205$ & 2021-01-03 & MIKE       & 9.03  & 600              & 248 & 2.5    & 0.98   & 0.97         \\
NYX122147 & 3521755401833022080 & 185.44804146 & $-15.12392540$ & 2021-07-28 & MIKE       & 11.48 & 1800             & 121 & 2.8    & 0.97   & 0.97         \\
NYX163705 & 1312596465842920576 & 249.27099451 & 31.32412562  & 2020-08-01 & HIRES      & 11.93 & 1000             & 144 & $-16.7$  & 1.00   & 0.82         \\
NYX175752 & 4609530533258729600 & 269.46600702 & 36.54835952  & 2020-08-01 & HIRES      & 10.65 & 300              & 155 & $-9.6$   & 0.97   & 0.95         \\
NYX175906 & 1345377378530252032 & 269.77384034 & 42.85162989  & 2020-08-01 & HIRES      & 12.33 & 1500             & 137 & $-7.6$   & 0.98   & 0.97         \\
NYX185206 & 2146631118469602944 & 283.02543696 & 53.80289724  & 2020-08-01 & HIRES      & 9.67  & 180              & 190 & $-0.9$   & 0.98   & 0.95         \\
NYX185250 & 2103594068696931968 & 283.21008746 & 40.77249320  & 2020-08-01 & HIRES      & 12.54 & 1800             & 149 & $-3.9$   & 0.97   & 0.87         \\
NYX195253 & 4195105698600078848 & 298.22200157 & $-9.45925775$  & 2020-08-01 & HIRES      & 12.46 & 1800             & 148 & $-5.6$   & 0.97   & 0.93         \\
NYX201634 & 2062121830140100992 & 304.14213600 & 39.18034350  & 2020-08-01 & HIRES      & 11.27 & 600              & 159 & 3.4    & 0.98   & 0.88         \\
NYX201856 & 2084248913225887104 & 304.73284567 & 48.34020559  & 2020-08-01 & HIRES      & 11.91 & 1000             & 143 & 4.7    & 0.97   & 0.97         \\
NYX204828 & 1751338782564204672 & 312.11838846 & 9.68257413   & 2020-08-01 & HIRES      & 10.14 & 300              & 179 & 3.6    & 0.97   & 0.88         \\
NYX205642 & 1757797833557540480 & 314.17519078 & 12.37187691  & 2020-08-01 & HIRES      & 12.04 & 1200             & 155 & 5.0      & 0.96   & 0.94         \\
NYX211145 & 1867283181371868544 & 317.93598233 & 35.27867523  & 2020-08-01 & HIRES      & 11.71 & 900              & 162 & 8.6    & 0.98   & 0.94         \\
NYX213012 & 1850722646463457920 & 322.55098693 & 31.10214042  & 2020-08-01 & HIRES      & 12.44 & 1800             & 148 & 10.3   & 0.99   & 0.93         \\
NYX214142 & 1797966051335964672 & 325.42542891 & 25.19027732  & 2020-08-01 & HIRES      & 11.39 & 600              & 142 & 11.2   & 0.98   & 0.97         \\
NYX215612 & 2695833495753971072 & 329.04880607 & 3.77654551   & 2020-08-01 & HIRES      & 12.81 & 2160             & 134 & 11.2   & 0.96   & 0.91         \\
NYX215744 & 1794930334091349632 & 329.43432692 & 23.58644105  & 2020-08-01 & HIRES      & 10.72 & 350              & 150 & 12.8   & 1.00   & 0.96         \\
NYX223035 & 2738050790890416384 & 337.64587000 & 18.37272779  & 2020-08-01 & HIRES      & 11.33 & 600              & 147 & 16     & 0.96   & 0.91         \\
NYX224245 & 2833632062543310848 & 340.68697508 & 19.16973836  & 2020-08-01 & HIRES      & 10.83 & 400              & 142 & 17.2   & 0.98   & 0.95         \\
NYX225510 & 2714860681911345280 & 343.79139075 & 9.53497180   & 2020-08-01 & HIRES      & 11.16 & 1200             & 215 & 17.9   & 0.95   & 0.90         \\
NYX230605 & 2815219469025808128 & 346.52177833 & 13.86879634  & 2020-08-01 & HIRES      & 11.20 & 500              & 140 & 19.3   & 0.97   & 0.91         \\
NYX230903 & 2385788396590080512 & 347.26428804 & $-23.09773201$ & 2020-08-01 & HIRES      & 11.67 & 900              & 150 & 14.1   & 0.97   & 0.87         \\
NYX232032 & 2813625898720075264 & 350.13478016 & 13.36493995  & 2020-08-01 & HIRES      & 9.05  & 80               & 153 & 20.5   & 0.96   & 0.93         \\
NYX234631 & 2771034730974538624 & 356.62832066 & 14.41544640  & 2020-08-01 & HIRES      & 12.58 & 1800             & 141 & 22.7   & 0.97   & 0.91
\enddata
\end{deluxetable*}
\end{longrotatetable}

\begin{deluxetable*}{lccccccccccccccc}
\tablecolumns{15}
\tabletypesize{\footnotesize}
\tablecaption{\label{tab:lines}Line Measurements}
\tablehead{Star & $\lambda$ (\AA) & ID & $\chi$ & $\log gf$ & EW & $\sigma$(EW) & FWHM (\AA) & $\log \epsilon_{i\,,\mathrm{ref}}$ & $d_i$ & $e_i$ 
& $\delta_{i,\teff}$ & $\delta_{i,\logg}$ & $\delta_{i,\nu_t}$ & $\delta_{i,\text{[M/H]}}$ 
}
\startdata
NYX201856   & 5778.45 &  26.0 &   2.59  & $ -3.43 $ &   9.5 &  0.4 &  0.11 & $ 6.92$ & $-0.35$ &  0.02 & $+0.05$ & $-0.00$ & $-0.00$ & $+0.01$ \\
NYX201856   & 5855.08 &  26.0 &   4.61  & $ -1.48 $ &   8.7 &  0.4 &  0.15 & $ 6.96$ & $-0.29$ &  0.02 & $+0.03$ & $+0.00$ & $-0.00$ & $+0.01$ \\
NYX201856   & 6027.05 &  26.0 &   4.08  & $ -1.09 $ &  40.5 &  0.3 &  0.13 & $ 7.00$ & $-0.31$ &  0.01 & $+0.04$ & $+0.00$ & $-0.02$ & $+0.02$ \\
NYX201856   & 6136.99 &  26.0 &   2.20  & $ -2.95 $ &  47.2 &  0.3 &  0.13 & $ 7.02$ & $-0.30$ &  0.01 & $+0.06$ & $+0.00$ & $-0.03$ & $+0.01$ \\
NYX201856   & 6151.62 &  26.0 &   2.18  & $ -3.29 $ &  30.3 &  0.3 &  0.12 & $ 6.97$ & $-0.31$ &  0.01 & $+0.06$ & $+0.00$ & $-0.01$ & $+0.01$ \\
NYX201856   & 6165.36 &  26.0 &   4.14  & $ -1.47 $ &  21.1 &  0.3 &  0.12 & $ 6.97$ & $-0.31$ &  0.01 & $+0.03$ & $+0.00$ & $-0.01$ & $+0.01$ \\
NYX201856   & 6173.33 &  26.0 &   2.22  & $ -2.88 $ &  47.4 &  0.4 &  0.12 & $ 6.98$ & $-0.30$ &  0.01 & $+0.06$ & $+0.00$ & $-0.03$ & $+0.01$ \\
NYX201856   & 6213.43 &  26.0 &   2.22  & $ -2.48 $ &  62.8 &  0.4 &  0.14 & $ 6.92$ & $-0.28$ &  0.01 & $+0.07$ & $+0.00$ & $-0.04$ & $+0.01$ \\
NYX201856   & 6271.28 &  26.0 &   3.33  & $ -2.70 $ &  11.7 &  0.4 &  0.15 & $ 6.94$ & $-0.24$ &  0.02 & $+0.04$ & $-0.00$ & $-0.00$ & $+0.01$ \\
NYX201856   & 6322.69 &  26.0 &   2.59  & $ -2.43 $ &  52.6 &  0.4 &  0.14 & $ 7.03$ & $-0.30$ &  0.01 & $+0.06$ & $+0.00$ & $-0.03$ & $+0.02$ \\
NYX201856   & 6335.33 &  26.0 &   2.20  & $ -2.18 $ &  76.8 &  0.4 &  0.15 & $ 6.87$ & $-0.25$ &  0.01 & $+0.08$ & $-0.00$ & $-0.05$ & $+0.01$ \\
NYX201856   & 6411.65 &  26.0 &   3.65  & $ -0.60 $ &  89.2 &  0.5 &  0.17 & $ 6.89$ & $-0.26$ &  0.01 & $+0.06$ & $-0.03$ & $-0.03$ & $+0.04$ \\
NYX201856   & 6518.37 &  26.0 &   2.83  & $ -2.44 $ &  37.7 &  0.4 &  0.14 & \nodata & \nodata &  0.01 & $+0.05$ & $+0.00$ & $-0.02$ & $+0.01$ \\
NYX201856   & 6581.21 &  26.0 &   1.49  & $ -4.68 $ &   9.2 &  0.4 &  0.13 & $ 7.01$ & $-0.42$ &  0.02 & $+0.06$ & $-0.00$ & $-0.00$ & $+0.01$ \\
NYX201856   & 6593.87 &  26.0 &   2.43  & $ -2.42 $ &  62.2 &  0.4 &  0.15 & $ 7.02$ & $-0.26$ &  0.01 & $+0.07$ & $+0.00$ & $-0.04$ & $+0.01$ \\
NYX201856   & 6609.11 &  26.0 &   2.56  & $ -2.69 $ &  41.4 &  0.4 &  0.14 & $ 7.01$ & $-0.32$ &  0.01 & $+0.06$ & $+0.00$ & $-0.02$ & $+0.01$ \\
NYX201856   & 6739.52 &  26.0 &   1.56  & $ -4.79 $ &   7.5 &  0.4 &  0.18 & $ 6.89$ & $-0.21$ &  0.03 & $+0.06$ & $-0.00$ & $-0.00$ & $+0.01$ \\
NYX201856   & 6810.26 &  26.0 &   4.61  & $ -0.99 $ &  22.1 &  0.4 &  0.15 & $ 6.98$ & $-0.34$ &  0.01 & $+0.03$ & $+0.00$ & $-0.01$ & $+0.02$ \\
NYX201856   & 6828.59 &  26.0 &   4.64  & $ -0.82 $ &  27.6 &  0.4 &  0.16 & $ 6.98$ & $-0.32$ &  0.01 & $+0.03$ & $+0.00$ & $-0.01$ & $+0.02$ \\
NYX201856   & 6837.01 &  26.0 &   4.59  & $ -1.69 $ &   7.2 &  0.4 &  0.18 & $ 7.01$ & $-0.26$ &  0.03 & $+0.03$ & $+0.00$ & $-0.00$ & $+0.01$ \\
NYX201856   & 6843.66 &  26.0 &   4.55  & $ -0.73 $ &  31.7 &  0.3 &  0.14 & $ 6.88$ & $-0.32$ &  0.01 & $+0.03$ & $+0.00$ & $-0.01$ & $+0.02$ \\
\enddata
\tablecomments{A portion of this table is shown for form. Column details are provided in the text. The full version is available online.}
\end{deluxetable*}

\subsection{Reference Stars for Differential Abundances}
\label{sec:method}
We adopt the method of differential abundance analysis, which minimizes systematic effects due to atomic data, stellar parameters, blended lines, and NLTE effects \citep[e.g.,][]{Nissen_2010, McWilliam_2013, Matsuno_2022}.\footnote{The coolest and most metal-rich Nyx185206 is removed from the differential abundance analysis due to blending issues affecting almost all elements of the star.}

Our Nyx stars span a large range of stellar parameters, so we adopt three different reference stars for our differential analysis: HIP7162, HIP88622 and Arcturus, which were analyzed in \citet{Bensby_2014} and \citet{McWilliam_2013}. We use HIP7162 as the reference star for all metal-poor stars ($\mbox{[Fe/H]} \sim -2.0$), with reference abundances from \citet{Bensby_2014}; we use HIP88622 as the reference star for metal-rich dwarfs ($\mbox{[Fe/H]} > -1.0$, $\teff \gtrsim 5200\,\unit{K}$), with reference abundances from \citet{Bensby_2014}; we use Arcturus as the reference star for the metal-rich giants ($\mbox{[Fe/H]} > -1.0$, $\teff \lesssim 5200\,\unit{K}$), with reference abundances from \citet{McWilliam_2013} and \citet{Fanelli_2021}. The adopted stellar parameters for these reference stars are in Table~\ref{tab:sp}.

\subsection{Stellar Parameters}
\label{sec:sp}
We derive stellar parameters spectroscopically, adopting ATLAS model atmospheres \citep{Castelli_2004}. We balance the excitation potential and line strength against differential Fe~I abundances to derive the effective temperature, $\teff$, and microturbulence, $\nu_t$, of the Nyx stars. 
We balance the ionization states for differential Fe~I and Fe~II to derive the surface gravity, $\log g$. The model metallicity of Nyx stars is set to differential $\mbox{[Fe~I/H]}$. We set $\mbox{\alphafe}$ to $\mbox{[Ca/Fe]}$. 
The differential line abundance of each element for each Nyx target is determined as follows: we match the individual lines and calculate the abundance differences between the target and the reference star. We add the reference abundance to the line abundance difference to derive the differential line abundance. The final differential abundance of each element is derived by adding the reference abundance to the average line abundance difference.

For the most metal-poor star Nyx010249, we are unable to effectively constrain its microturbulence, so we determine it by fitting a polynomial to the $\log g-\nu_t$ relation of the other Nyx stars, $\nu_t = 0.053 (\log g)^2 -0.47 \log g + 2.08$.

Stellar parameter uncertainties include a combination of statistical and systematic uncertainties.
Statistical uncertainties are determined for the $1 \sigma$ standard error on the slopes of abundance differences (see \citealt{Ji_2020} for details).
Systematic uncertainties are estimated by comparing to photometric stellar parameters using \Gaia~DR3 broad-band photometry \citep{Gaia_DR3}. 
The photometric effective temperature of each star is determined using the color$-\teff$ relation \citep{Mucciarelli_2021}. The microturbulence and model metallicity are determined using \code{SMHR}, and are individually examined to ensure accuracy. The final systematic uncertainties are $50\,\unit{K}$, 0.15~dex, $0.10\kms$ and 0.05~dex for $\teff$, $\log g$, $\nu_t$ and $\mbox{[M/H]}$ respectively for metal-rich dwarfs; $100\,\unit{K}$, 0.30~dex, $0.10\kms$ and 0.20~dex for metal-rich giants; and $200\,\unit{K}$, 0.40~dex, $0.10\kms$ and 0.20~dex for metal-poor stars.

\begin{figure}
\begin{center}
\includegraphics[width=0.95\textwidth]{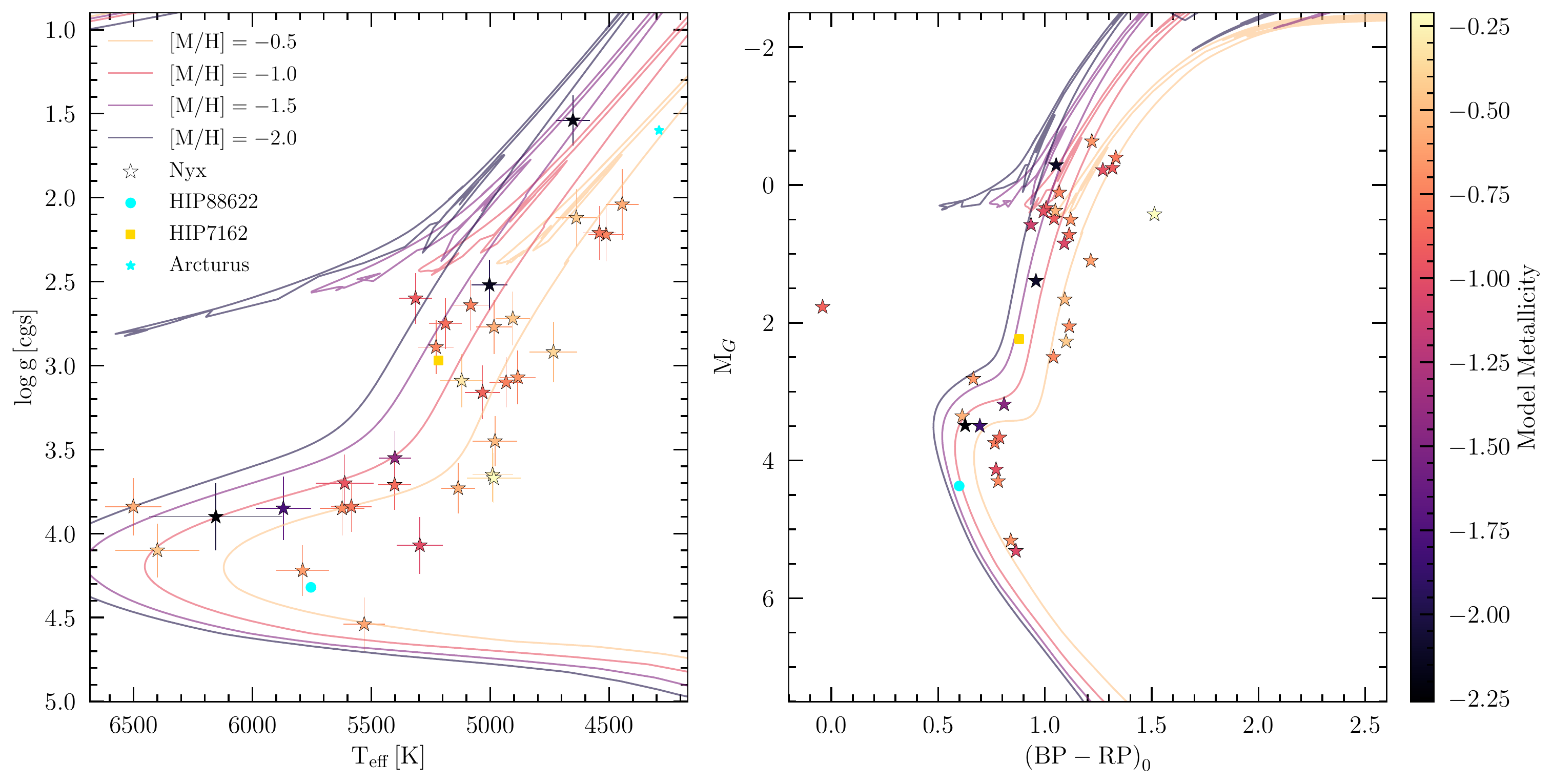} \\
\end{center}
\caption{Kiel diagram (left) and Color Magnitude diagram (right) of the observed Nyx stars. Nyx stars are color-coded by model metallicity. MIST isochrones with age 10~Gyr and $\mbox{[M/H]} \in [-2.0, -0.5]$ \citep{2016ApJ...823..102C} are plotted for comparison. The reference stars HIP7162, HIP88622 and Arcturus used in Sec.~\ref{sec:method} are indicated by the gold square, cyan dot and star, respectively. 
The plot demonstrates that the Nyx sample is composed of slightly metal-poor dwarf and giant stars, with five Nyx stars having distinctly lower model metallicity.
}
\label{fig:stellar_parameters}
\end{figure}

Table~\ref{tab:sp} shows the stellar parameters of the Nyx and reference stars, HIP88622, HIP7162 and Arcturus. Each row provides a star's $\teff$, $\logg$, $\nu_t$, the model metallicity ($\mbox{[M/H]}$), $\alpha$ abundance ($\mbox{\alphafe}$) and the differential Fe abundance ($\text{[Fe/H]}\textsubscript{d}$).\footnote{The differential Fe abundance $\text{[Fe/H]\textsubscript{d}}$ is shown to contrast with the model metallicity $\mbox{[M/H]}$ derived by balancing iron abundances, and will henceforth be referred to as the metallicity exclusively.} Uncertainties $\sigma(X)$ are provided for the first four of these quantities; uncertainties on $\mbox{\alphafe}$ are not propagated but are negligible. 
Fig.~\ref{fig:stellar_parameters} shows the Kiel and \Gaia~DR3 Color-Magnitude diagrams of the Nyx and two of the reference stars.\footnote{The reference star Arcturus is not in Gaia~DR3.} In the Color-Magnitude Diagram, reddening is corrected using $E(B-V)$ from \citet{Schlafly_2011}\footnote{Obtained from the IPAC dust map \citep{IRSA}.} and transformed to $G$, $BP$, and $RP$ magnitudes using the \Gaia~DR3 extinction coefficients\footnote{https://www.cosmos.esa.int/web/gaia/edr3-extinction-law}.
To guide the eye, we show MIST isochrones \citep{2016ApJ...823..102C} with the age set at 10~Gyr and covering the metallicity range $[-2.0, -0.5]$. 
The plot demonstrates that the sample stars are composed of slightly metal-poor dwarf and giant stars. The metallicity of most Nyx stars falls in the range $\mbox{[M/H]} \in [-1.0, -0.5]$, but there are five stars with metallicity $\mbox{[M/H]} \sim -2.0$. 
Compared to the Nyx targets in \citet{2020NatAs...4.1078N}, those in this work contain more metal-rich giants.

\begin{deluxetable*}{lcccccccccc}
\tablecolumns{11}
\tabletypesize{\footnotesize}
\tablecaption{\label{tab:sp}Stellar Parameters}
\tablehead{ Star & $\teff (\unit{K})$ & $\sigma(\teff)$ & $\log g$ (dex) & $\sigma(\log g)$ & $\nu_t (\kms)$ & $\sigma(\nu_t)$ & $\mbox{[M/H]}$ & $\sigma(\mbox{[M/H]})$ & $\mbox{\alphafe}$ & $\text{[Fe/H]}\textsubscript{d}$ }
\startdata
NYX004259   & 6400 &  177 &  4.10 &  0.16 &  1.60 &  0.23 & $-0.45$ &  0.11 &  0.30 & $-0.30$ \\
NYX010249   & 6154 &  280 &  3.90 &  0.20 &  1.01 &  1.05 & $-2.26$ &  0.14 &  0.40 & $-2.20$ \\
NYX010842   & 4980 &   94 &  3.45 &  0.15 &  0.89 &  0.11 & $-0.50$ &  0.13 &  0.40 & $-0.42$ \\
NYX013214   & 5296 &   97 &  4.07 &  0.17 &  0.82 &  0.24 & $-1.01$ &  0.13 &  0.40 & $-0.86$ \\
NYX014107   & 5082 &   77 &  2.64 &  0.15 &  1.48 &  0.09 & $-0.75$ &  0.12 &  0.40 & $-0.68$ \\
NYX014725   & 5120 &   90 &  3.09 &  0.16 &  1.19 &  0.09 & $-0.31$ &  0.12 &  0.40 & $-0.22$ \\
NYX022545   & 4734 &  100 &  2.92 &  0.18 &  1.16 &  0.10 & $-0.39$ &  0.13 &  0.40 & $-0.32$ \\
NYX035539   & 4513 &   74 &  2.22 &  0.16 &  1.43 &  0.09 & $-0.74$ &  0.12 &  0.40 & $-0.67$ \\
NYX044831   & 5314 &   69 &  2.60 &  0.15 &  1.70 &  0.11 & $-0.93$ &  0.11 &  0.40 & $-0.86$ \\
NYX063721   & 4984 &   76 &  2.77 &  0.16 &  1.38 &  0.09 & $-0.56$ &  0.12 &  0.40 & $-0.47$ \\
NYX064943   & 4638 &   87 &  2.12 &  0.17 &  1.50 &  0.09 & $-0.44$ &  0.13 &  0.40 & $-0.36$ \\
NYX122147   & 4652 &   71 &  1.54 &  0.15 &  1.32 &  0.15 & $-2.19$ &  0.13 &  0.40 & $-2.13$ \\
NYX163705   & 5530 &   88 &  4.54 &  0.16 &  0.85 &  0.20 & $-0.60$ &  0.12 &  0.40 & $-0.47$ \\
NYX175752   & 4445 &   69 &  2.04 &  0.21 &  1.61 &  0.09 & $-0.57$ &  0.12 &  0.40 & $-0.49$ \\
NYX175906   & 5870 &  117 &  3.85 &  0.19 &  0.85 &  0.64 & $-1.78$ &  0.14 &  0.40 & $-1.71$ \\
NYX185250   & 5228 &   75 &  2.89 &  0.16 &  1.59 &  0.10 & $-0.74$ &  0.12 &  0.40 & $-0.66$ \\
NYX195253   & 5623 &   93 &  3.85 &  0.16 &  1.12 &  0.14 & $-0.70$ &  0.12 &  0.40 & $-0.56$ \\
NYX201634   & 4933 &   70 &  3.10 &  0.15 &  1.08 &  0.08 & $-0.75$ &  0.12 &  0.40 & $-0.67$ \\
NYX201856   & 5402 &   68 &  3.71 &  0.15 &  0.90 &  0.11 & $-0.84$ &  0.11 &  0.40 & $-0.70$ \\
NYX204828   & 4541 &   70 &  2.21 &  0.16 &  1.37 &  0.09 & $-0.80$ &  0.12 &  0.40 & $-0.71$ \\
NYX205642   & 5135 &   71 &  3.73 &  0.15 &  1.07 &  0.09 & $-0.55$ &  0.11 &  0.40 & $-0.47$ \\
NYX211145   & 4884 &   75 &  3.07 &  0.16 &  1.27 &  0.09 & $-0.66$ &  0.12 &  0.40 & $-0.58$ \\
NYX213012   & 5003 &   76 &  2.52 &  0.15 &  0.91 &  0.17 & $-2.14$ &  0.12 &  0.40 & $-2.08$ \\
NYX214142   & 5188 &   69 &  2.75 &  0.15 &  1.49 &  0.09 & $-0.83$ &  0.11 &  0.40 & $-0.74$ \\
NYX215612   & 5789 &  111 &  4.22 &  0.15 &  1.26 &  0.21 & $-0.63$ &  0.12 &  0.40 & $-0.51$ \\
NYX215744   & 4990 &   85 &  3.65 &  0.16 &  0.92 &  0.12 & $-0.28$ &  0.12 &  0.40 & $-0.20$ \\
NYX223035   & 4905 &   73 &  2.72 &  0.16 &  1.31 &  0.09 & $-0.45$ &  0.12 &  0.40 & $-0.38$ \\
NYX224245   & 4985 &  114 &  3.67 &  0.15 &  0.80 &  0.14 & $-0.21$ &  0.13 &  0.40 & $-0.13$ \\
NYX225510   & 5584 &   86 &  3.84 &  0.15 &  1.10 &  0.13 & $-0.79$ &  0.11 &  0.40 & $-0.66$ \\
NYX230605   & 6501 &  118 &  3.84 &  0.17 &  1.40 &  0.19 & $-0.56$ &  0.11 &  0.25 & $-0.42$ \\
NYX230903   & 5401 &   68 &  3.55 &  0.16 &  0.96 &  0.11 & $-1.41$ &  0.11 &  0.30 & $-1.35$ \\
NYX232032   & 5032 &   74 &  3.16 &  0.16 &  1.14 &  0.09 & $-0.88$ &  0.12 &  0.40 & $-0.80$ \\
NYX234631   & 5612 &  121 &  3.70 &  0.17 &  0.79 &  0.28 & $-0.96$ &  0.13 &  0.40 & $-0.82$ \\
Arcturus    & 4290 &  119 &  1.60 &  0.31 &  1.60 &  0.22 & $-0.50$ &  0.14 &  0.40 & $-0.49$ \\
HIP7162     & 5217 &   92 &  2.97 &  0.16 &  1.60 &  0.49 & $-2.05$ &  0.12 &  0.40 & $-2.05$ \\
HIP88622    & 5754 &   80 &  4.32 &  0.15 &  1.00 &  0.13 & $-0.53$ &  0.11 &  0.40 & $-0.40$ \\
\enddata
\tablecomments{Stellar parameters of the Nyx stars. $\text{{[X/Fe]}d}$ refers to the differential Fe abundance, which differs from the model metallicity $\text{{[M/H]}}$. See Sec.~\ref{sec:method} for more details.}
\end{deluxetable*}

\begin{deluxetable*}{llcccccccccc}
\tablecolumns{12}
\tabletypesize{\footnotesize}
\tablecaption{\label{tab:abunds}Stellar Abundances}
\setlength{\tabcolsep}{3pt}
\tablehead{Star & El. & $N$ & 
$\text{[X/H]}\textsubscript{d}$ &
$\text{{[X/Fe]}}\textsubscript{d}$ & $\sigma_{\text{{[X/Fe]}d}}$ &
$\text{[X/H]}\textsubscript{ref}$ &
$\sigma_{\text{{[X/H]}}}\textsubscript{ref}$ &
$\text{[X/Fe]}\textsubscript{ref}$ &
$\sigma_{\mathrm{rand}}$ &
$\sigma_{\mathrm{sys}}$ &
$\sigma_{\mathrm{sp}}$
}
\startdata
NYX201856       & Na I  &   4 & $-0.68$ & $ 0.02$ &  0.06 &$-0.37$ &  0.02 & $ 0.03$ & 0.00 &  0.04 &  0.04 \\
NYX201856       & Mg I  &   4 & $-0.34$ & $ 0.36$ &  0.06 &$-0.15$ &  0.02 & $ 0.25$ & 0.00 &  0.02 &  0.06 \\
NYX201856       & Si I  &   6 & $-0.55$ & $ 0.14$ &  0.12 &$-0.23$ &  0.03 & $ 0.17$ & 0.00 &  0.12 &  0.03 \\
NYX201856       & Ca I  &   7 & $-0.46$ & $ 0.24$ &  0.07 &$-0.29$ &  0.02 & $ 0.11$ & 0.00 &  0.02 &  0.07 \\
NYX201856       & Ti I  &  25 & $-0.37$ & $ 0.33$ &  0.09 &$-0.19$ &  0.01 & $ 0.21$ & 0.00 &  0.05 &  0.07 \\
NYX201856       & Fe I  &  20 & $-0.70$ & $ 0.00$ &  0.07 &$-0.40$ &  0.01 & $ 0.00$ & 0.00 &  0.04 &  0.06 \\
NYX201856       & Fe II &   5 & \nodata & \nodata &  0.11 &\nodata & \nodata & \nodata & 0.01 &  0.05 &  0.09 \\
NYX201856       & Ni I  &  12 & $-0.69$ & $ 0.01$ &  0.07 &$-0.40$ &  0.01 & $ 0.00$ & 0.00 &  0.04 &  0.06 \\
NYX201856       & Zn I  &   2 & $-0.53$ & $ 0.17$ &  0.08 &$-0.22$ &  0.02 & $ 0.18$ & 0.01 &  0.02 &  0.08 \\
NYX201856       & Y II  &   3 & $-0.66$ & $ 0.04$ &  0.10 &$-0.56$ &  0.03 & $-0.16$ & 0.01 &  0.05 &  0.09 \\
NYX201856       & Ba II &   3 & $-0.68$ & $ 0.02$ &  0.10 &$-0.47$ &  0.04 & $-0.07$ & 0.00 &  0.04 &  0.10 \\
\enddata
\tablecomments{One star from this table is shown for form. Column details are provided in the text. The full version is available online.
}
\end{deluxetable*}

\subsection{Abundance Uncertainties}
\label{sec:diff}

Differential abundance uncertainties were calculated using the method introduced in \citet{McWilliam_2013}. The abundance uncertainty for each species is 
\begin{align}\label{eq:error}
\begin{split}
    {\sigma}^2 &= {\sigma^2_{\mathrm{rand}}} + {\sigma^2_{\mathrm{sys}}} + {\sigma^2_{\mathrm{sp}}}
    \\
    &=
    \frac{1}{\sum_{i}^{N} 1/{e_i}^2} + \frac{1}{N} \sum_{i}^{N} {(d_i - \langle d \rangle)}^2 + \sum_X  \left(\frac{1}{N} \sum_{i}^{N} \delta_{i\,, X}\right)^2 \,,
\end{split}
\end{align}    
where $i$ is the index of a line for the species, $N$ is the total number of lines for the species, $e_i$ is the statistical uncertainty, $d_i$ is the line-by-line abundance difference relative to the reference star, $\langle d \rangle$ is the mean line-by-line abundance difference, $X$ is the stellar parameter ($\teff$, $\log{g}$, $\nu_t$ and $\mbox{[Fe/H]}$), and $\delta_{i\,, X}$ is the stellar parameter uncertainty difference (see Table~\ref{tab:lines}).
The individual terms of Eqn.~\ref{eq:error} are provided in Table~\ref{tab:abunds}.
The first term on the right hand side of the equation is the squared average uncertainty due to random spectrum noise; the second term is the squared average systematic uncertainty as measured by the line-to-line standard deviation; and the third term is the squared uncertainty due to stellar parameters. We regard the precision of our abundance measurements to be  $\sqrt{{\sigma^2_{\mathrm{rand}}}+{\sigma^2_{\mathrm{sys}}}}$ ($\sigma_{\rm [X/Fe]d}$ in Table~\ref{tab:abunds}). 
The accuracy of our abundances should further add the stellar parameter uncertainty in quadrature.
Note that the covariance between stellar parameters is included in \citet{McWilliam_2013} but not considered in this paper, which tends to overestimate the abundance uncertainties for $\mbox{[X/Fe]}$ ratios (see \citealt{Ji_2020} for details).

Table~\ref{tab:lines} lists the individual measurements for each line in our line list. 
As an example, the portion of the table printed here lists the Fe~I line measurements of Nyx201856.
Each row contains the name of the star, the wavelength of the line~($\lambda$), the element species~(ID), the excitation potential~($\chi$) and oscillator strength~($\log gf$), the equivalent width and uncertainty~(EW, $\sigma$(EW)), the full-width-half-max~(FWHM), the measured abundance of the reference star~($\log \epsilon_{i\,,\mathrm{ref}}$), the abundance difference~($d_i$), the statistical uncertainty~($e_i$) and the stellar parameter abundance differences~($\delta_{i\,,X}$), where $X$ is a stellar parameter.

The differential abundances and uncertainties are given in Table~\ref{tab:abunds}. The differential abundances of Nyx201856 are measured with respect to the reference star HIP88622. 
In the table, each row contains the name of the star; the element measured~(El.); the number of lines used~(N); 
the derived differential abundances $\mbox{[X/H]}$ \citep[relative to the solar abundance;][]{Asplund_2009} and $\mbox{[X/Fe]}$ with respect to HIP88622~($\mathrm{[X/H]}\textsubscript{d}$, $\mathrm{[X/Fe]}\textsubscript{d}$), and the uncertainty $\sigma_{\mathrm{[X/Fe]d}}$; 
the $\mbox{[X/H]}$ value for the reference star and uncertainty~($\mathrm{[X/H]}\textsubscript{ref}$, $\sigma_{\mathrm{{[X/H]}}}\textsubscript{ref}$); the $\mbox{[X/Fe]}$ value for the reference star~($\mathrm{[X/Fe]}\textsubscript{ref}$); uncertainty due to the random spectrum noise $\sigma_{\mathrm{rand}}$, systematic uncertainty $\sigma_{\mathrm{sys}}$, and uncertainty due to stellar parameters $\sigma_{\mathrm{sp}}$.

We used spectroscopic stellar parameters to derive the abundances, but for verification, we also determined the differential abundances of the Nyx stars using photometric stellar parameters. For most elements, the differential abundances measured with photometric stellar parameters have a slightly larger standard deviation ($\sim0.02$~dex) than those measured with spectroscopic stellar parameters. 
The conclusions in Sec.~\ref{sec:conclusion} remain the same regardless of the use of spectroscopic or photometric stellar parameters, since any systematic stellar parameter uncertainties should be accounted for in our error analysis.

We also determined absolute abundances, which allowed us to measure a few more extra elements not measured in \citet{Bensby_2014}, see Appendix~\ref{sec:abs}.

\subsection{Discussion of Individual Element Abundances and Lines}
\label{sec:elems}

We now discuss the selection of lines for elements included in the differential abundance analysis. For metal-rich giants, the selected lines are checked for blending issues using visual inspection (Magnesium and Titanium) or by comparing against spectral synthesis (Yttrium). The selected lines are examined for saturation using their reduced equivalent widths (REW) and by visual inspection. Saturated lines are fitted with Voigt profiles.

\textit{\textul{Iron.}} 
We measure up to 29  Fe~I and Fe~II lines in most Nyx stars.
Fe~I lines are used to determine the stellar parameters and model metallicities of the Nyx stars. Fewer Fe lines from the \citet{Jonsson_2017} line list could be measured for the metal-poor ($\mbox{[Fe/H]}\sim -2.0$) stars. Therefore, extra lines from \citet{Bensby_2005} are added to enable a stellar parameter and metallicity determination. 
Since we balance the ionization states to determine $\logg$, the $\mbox{[Fe~I/H]}$ and $\mbox{[Fe~II/H]}$ we measure usually differ by less than 0.05~dex.

\textit{\textul{Sodium.}} 
Na~I is measured using up to four lines, including the Na doublet at 6154~\AA~and 6160~\AA.

\textit{\textul{Magnesium.}} 
Equivalent widths of up to six Mg~I lines, including the lines at 6318 and 6319~\AA,\footnote{We fit the local continuum around the Mg lines at 6318 and 6319~\AA, including the depression from the Ca~I auto-ionization feature.} are measured in most Nyx stars.
The Mg line at 4702~\AA~is usually saturated.

\textit{\textul{Silicon.}} 
Equivalent widths of six Si~I lines are measured in most Nyx stars.

\textit{\textul{Calcium.}} 
Ca has the smallest uncertainties out of the $\alpha$-elements and is used as the reference $\alpha$ element in determining the stellar parameters of the Nyx stars. Equivalent widths of seven Ca~I lines are measured in a typical Nyx star. Ca lines at 6122~\AA, 6162~\AA, and 6439~\AA~are usually saturated.

\textit{\textul{Titanium.}}
We determine Ti~I abundances by equivalent width measurements of up to 25 Ti~I lines.

\textit{\textul{Nickel.}} 
We measure the equivalent widths of up to 12 Ni~I lines. Since Ni abundances were not determined in \citet{McWilliam_2013}, the reference Ni abundance of Arcturus is instead taken from \citet{Fanelli_2021}. We add an uncertainty of 0.1~dex (approximately the difference between the $\mbox{[Fe/H]}$ of Arcturus in \citet{McWilliam_2013} and that in \citet{Fanelli_2021}) to the final Ni abundances to account for the offset (the ``zero-point uncertainty", see Fig.~\ref{chemical_abundances1}).

\textit{\textul{Zinc.}} 
Two Zn lines are measured at 4722~\AA~and 4810~\AA~using equivalent widths. 
Since Zn is not measured for the reference star HIP7162, we are unable to determine the differential abundances of Zn for the metal-poor Nyx stars.
The reference Zn abundance of Arcturus is taken from \citet{Fanelli_2021}, and an uncertainty of 0.1~dex is added to the final Zn abundances to account for the zero-point uncertainty.

\textit{\textul{Yttrium.}} 
In general, up to six Y~II lines are measured using equivalent widths.
For the very cool metal-rich giant reference star Arcturus ($\mbox{[Fe/H]} = -0.49$, $\teff = 4290\,\unit{K}$), only the Y lines at longer wavelengths ($\lambda \gtrsim 5200~\r{A}$) could be reliably measured.

\textit{\textul{Barium.}} Abundances of three Ba~II lines are measured using equivalent widths. We verify that including hyperfine splitting does not significantly affect the Ba abundances.

\textit{\textul{Lanthanum.}} 
La~II abundances are derived by synthesizing the line at 6391~\AA, which has hyperfine splitting and is only available in the metal-rich Nyx stars. For metal-rich giants, the differential La abundances are measured with respect to Arcturus, whose reference abundance is from \citet{McWilliam_2013}. Although \citet{Bensby_2014} did not determine the La abundances, the same group measured the La abundances in \citet{Battistini_2016} for a subset of their thick disk stars.
Unfortunately, \citet{Battistini_2016} did not measure La for  our reference star HIP88622. To estimate the differential La abundance for metal-rich dwarfs, we instead assume that the La abundance of HIP88622 is identical to HIP16365, which has nearly identical stellar parameters. The La abundance of HIP16365 is measured using the different 4662~{\AA} line, so we estimate an uncertainty of HIP88622 (see Fig.~\ref{chemical_abundances1})
using the standard deviation of the La abundance measurements for the 4662~{\AA} line in \citet{Battistini_2016}.

\section{Abundance Results}
\label{sec:result}
\begin{figure*}
\begin{center}
\includegraphics[width=0.8\textwidth]{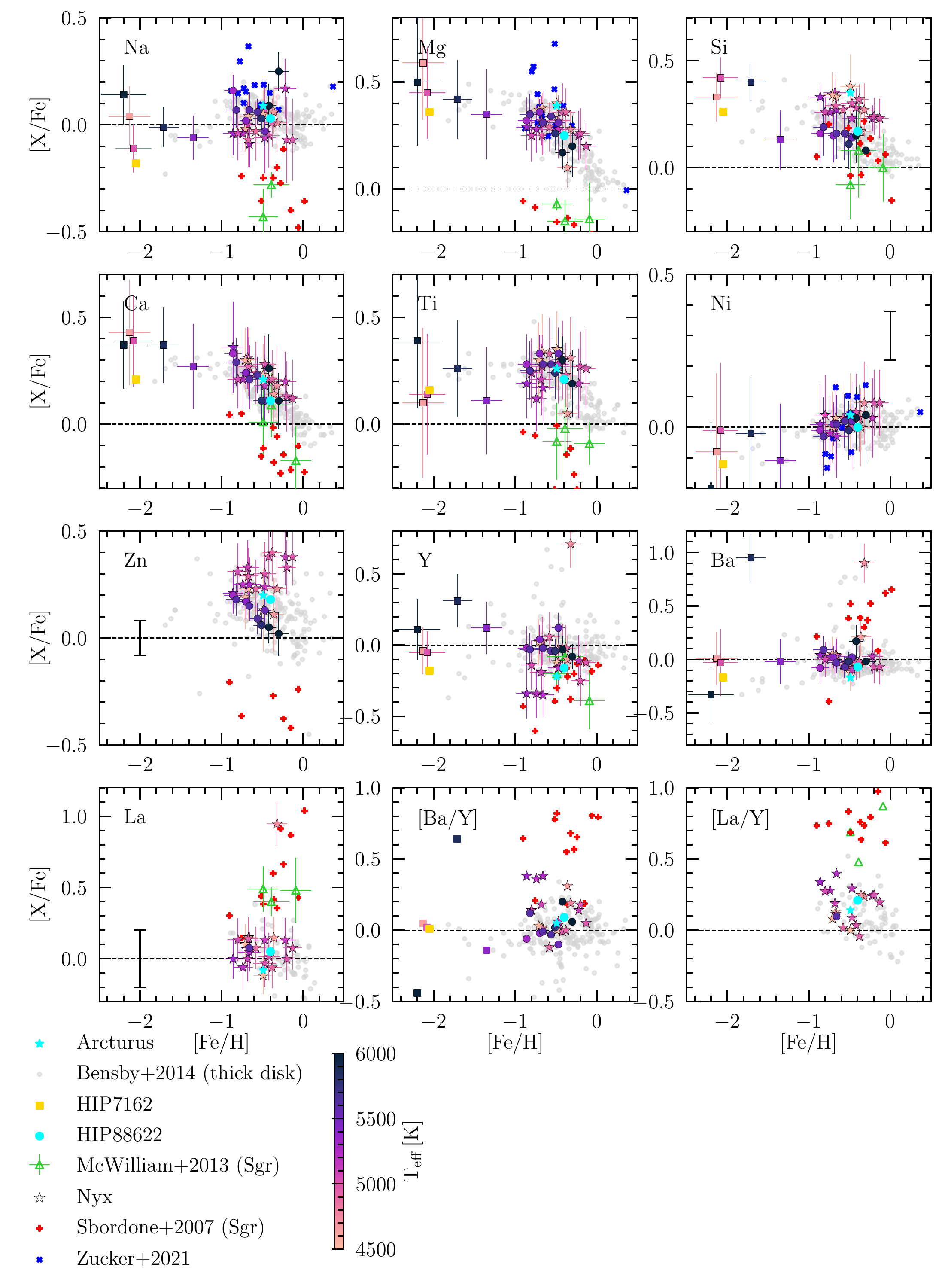} \\
\end{center}
\caption{Differential chemical abundances ($\mbox{[X/Fe]}$ versus $\mbox{[Fe/H]}$, y-axis is $\mbox{[X/Fe]}$ except when another ratio is indicated) of metal-rich Nyx dwarfs (dots) relative to HIP88622 (cyan dot), giants (stars, labelled as ``Nyx") relative to Arcturus (cyan star), and metal-poor stars (squares) relative to HIP7162 (gold square), color-coded by effective temperature. 
For Na, Mg, and Ni, 18 Nyx stars selected by \citet{Zucker_2021} from GALAH~DR3 are plotted as blue crosses. The abundances of thick disk stars (light gray dots) are from \citet{Bensby_2014}, except for La abundances, which are from \citet{Battistini_2016}.
The abundances of three Sgr dwarf galaxy stars from \citet{McWilliam_2013} are indicated by the green triangles, while those of 12 Sgr stars from \citet{Sbordone_2007} are indicated by the red plusses. A dashed line at $\mbox{[X/Fe]} = 0.0$ is shown as a reference. The black errorbar in the Ni, Zn and La abundance panel indicates the zero-point uncertainty. In general, the chemical abundances of the metal-rich Nyx component are consistent with the thick disk and distinct from the Sgr dwarf galaxy. 
}
\label{chemical_abundances1}
\end{figure*}

Table~\ref{tab:abunds} presents the final abundances of each Nyx star. Fig.~\ref{chemical_abundances1} shows the abundance comparison of Nyx stars with thick disk stars from \citet{Bensby_2014}. We highlight the reference stars Arcturus, HIP88622 and HIP7162, which are used for the differential abundance analysis of the 19 metal-rich giants ($\mbox{[Fe/H]}>-1.0$, $\teff \lesssim 5200\,\unit{K}$), nine metal-rich dwarfs ($\mbox{[Fe/H]}>-1.0$, $\teff \gtrsim 5200\,\unit{K}$) and five metal-poor ($\mbox{[Fe/H]}\sim-2.0$) Nyx stars, respectively. The stars are color-coded by $\teff$ to highlight possible systematic issues. 
Fig.~\ref{chemical_abundances1} also shows the abundances of Nyx stars from GALAH~DR3~\citep{Zucker_2021}, which we note are not on exactly the same differential abundance scale.
We use chemical abundances of three Sagittarius (Sgr) dwarf galaxy stars from \citet{McWilliam_2013} and 12 Sgr stars from \citet{Sbordone_2007} to illustrate the expected abundance differences between metal-rich dwarf galaxy stars and thick disk stars.

A quick glance at Fig.~\ref{chemical_abundances1} shows that most Nyx stars are relatively metal-rich ($\mbox{[Fe/H]}\sim -0.5$), but five stars have $\mbox{[Fe/H]} \sim -2.0$. Detailed chemical abundances of these metal-poor Nyx stars were not previously discussed \citep{2020NatAs...4.1078N,Zucker_2021,Horta_2022}. In what follows, we will discuss the chemical abundances of the metal-rich component and the metal-poor component of Nyx separately.

\subsection{Metal-Rich Component}

Overall, the metal-rich Nyx component is consistent with the thick disk and distinct from dwarf galaxies. Specifically, Nyx stars have high $\alpha$ abundances and do not show evidence of metal-poor s-process.

\textul{\textit{Odd-Z element (Na).}}
The metal-rich Nyx stars average $\mbox{[Na/Fe]}=0.02$, with a dispersion of 0.08~dex. Fig.~\ref{chemical_abundances1} demonstrates that the Na abundances of the metal-rich Nyx stars are consistent with the thick disk stars within $\sim 0.1$~dex.
The Na abundances are also much higher (by $\sim 0.3$~dex) than the Sgr dwarf galaxy stars.

\textul{\textit{$\alpha$-elements (Mg, Si, Ca, Ti).}}
Fig.~\ref{chemical_abundances1} shows that the metal-rich Nyx stars have high $\alpha$ abundances with a slightly downward trend with $\mbox{[Fe/H]}$, which is consistent with the high-$\alpha$ thick disk. The $\alpha$ abundances of the metal-rich Nyx stars are inconsistent with metal-rich dwarf galaxies---typically $0.2-0.4$~dex higher than those of the Sgr stars. There may appear to be an overlap between the $\alpha$ abundances of Nyx and Sgr stars at the highest metallicity, yet in this region, the most metal-rich thick disk stars have similar $\alpha$ abundances as well.
Fig.~\ref{chemical_abundances1} illustrates that the Mg abundances of the metal-rich Nyx component has a plateau at $\mbox{[Fe/H]}\lesssim-0.5$ and a slightly downward trend at higher metallicities. This agrees with the Nyx abundances reported in \citet{Zucker_2021} and \citet{Horta_2022}, but differs from the RAVE-on Mg abundances reported in \citet{2020NatAs...4.1078N} which were lower and did not show such a trend. \citet{Zucker_2021} pointed out that this is likely due to low precision in the RAVE-on Mg abundance measurements. 
For the metal-rich giants, the large uncertainties in the Ti abundances are mainly systematic uncertainties ($\delta_{\teff} \sim 0.17$) resulting from the large $\teff$ uncertainties (on average, $\sim 100\,\unit{K}$).

\textul{\textit{Iron-peak elements (Ni and Zn).}}
The iron-peak abundances have a relatively small dispersion of 0.03~dex (Ni) and 0.1~dex (Zn). In Fig.~\ref{chemical_abundances1}, the $\mbox{[X/Fe]}$ ratios of iron-peak elements do not display significant trends with respect to $\mbox{[Fe/H]}$. The Ni abundances from GALAH~DR3 \citep{Zucker_2021} have a larger scatter than our differential abundances for Ni, but the overall trend remains flat. Fig.~\ref{chemical_abundances1} shows relatively large uncertainties in the Ni and Zn abundance measurements. These are mainly systematic uncertainties ($\delta_{\log g} \sim 0.15$) resulting from the large $\log g$ uncertainties ($\sim 0.3$~dex).
In general, the abundances of iron-peak elements for the metal-rich Nyx stars are all consistent with the thick disk stars.

\textul{\textit{Neutron-capture elements (Y, Ba, La, Eu).}}
Most metal-rich Nyx stars have $\mbox{[Y/Fe]}$ and $\mbox{[Ba/Fe]}$ $\sim 0.0$,  similar to the thick disk, although there appears to be a substantial scatter around this mean (0.19~dex for Y and 0.18~dex for Ba).
In Fig.~\ref{chemical_abundances1}, the average Y abundance of the metal-rich dwarfs are higher by $\sim 0.3$~dex when compared to the Sgr dwarf galaxy stars.

Some of the Ba scatter might be expected from inhomogeneous s-process enrichment or binary mass transfer from an Asymptotic Giant Branch (AGB) companion \citep[e.g.,][]{Cseh_2022}. In order to evaluate the binary nature of the Nyx stars, we compare our observed radial velocities with \Gaia~DR3 results \citep{Gaia_DR3}. We find three Nyx stars with radial velocity variations more than $4.2\,\kms$ ($3\,\sigma$). One of these stars is the metal-rich star Nyx022545, which also has a very high Ba abundance ($\mbox{[Ba/Fe]} \sim 1.0$.) These stars could potentially be in close binaries \citep[e.g.,][]{Monaco_2007}. The metal-poor star Nyx175906 with a relatively high $\mbox{[Ba/Fe]}$, however, has an observed radial velocity consistent with \Gaia~DR3, which cannot determine its binary nature.
The $\mbox{[Ba/Y]}$ ratios of the metal-rich component also broadly match the thick disk, in contrast to dwarf galaxies that are expected to have higher $\mbox{[Ba/Y]}$ ratios \citep[$\sim0.7$;][]{Venn_2004,McWilliam_2013}.

Fig.~\ref{chemical_abundances1} indicates that the La abundances of the Nyx stars are roughly constant as a function of metallicity, and are $\gtrsim 0.4$~dex lower than those of the Sgr dwarf galaxy stars. The $\mbox{[La/Y]}$ ratios of the Sgr stars are also much higher, characteristic of the ratio of heavy to light neutron-capture elements of nearby dwarf galaxies \citep[$0.5-0.8$;][]{Shetrone_2001,Shetrone_2003}. A few metal-rich Nyx giants have $\mbox{[La/Y]}\gtrsim 0.5$. Unlike the Sgr stars, this is due to their low Y abundance, not high La abundance.

We measured Eu in five Nyx stars observed with MIKE. See Appendix~\ref{sec:abs} for more details.

\subsection{Metal-Poor Tail}

We first discuss the metal-poor stars of the Bensby sample, against which we will compare the abundances of the Nyx stars. The average $\mbox{[Fe/H]}$ of thick disk stars from \citet{Bensby_2014} is likely biased, since the selection function for the thick disk used by \citet{Bensby_2014} was complex and could not be used to reliably determine the metallicity distribution.

The most metal poor stars in the Bensby sample have $-1.8 \leq \mbox{[Fe/H]}\sim -1$. The probabilities of these stars belonging to the thick disk relative to the halo (see definition of thick disk to halo ratio in Sec.~\ref{sec:obs}) is between $\rm{TD/H}\sim15$ and $\rm{TD/H}\sim1800$, making them unlikely to be halo stars contaminating the sample.

The Nyx stream has a metal-poor tail that consists of five stars with $\mbox{[Fe/H]}\sim -2.0$ (compared to the average thick disk star from \citet{Bensby_2014} with $\mbox{[Fe/H]}\sim -0.4)$. Although the sample size of this subset is small, these are clearly isolated in metallicity from the metal-rich majority of the Nyx stars. More specifically, the average metallicity of the metal-poor Nyx stars is $-2.19^{+0.47}_{-0.07}$, compared to
$-0.56^{+0.22}_{-0.25}$ of the metal-rich Nyx stars, where the lower and upper limits are the 16th and 84th percentiles respectively.

The differential abundances of Nyx stars in the metal-poor tail are derived with respect to the similarly metal-poor reference star HIP7162. Fig.~\ref{chemical_abundances1} shows that the differential abundances of the five most metal-poor stars are broadly similar to (within 0.2~dex of) those of the more metal-poor outliers in the Bensby thick disk stars \citep{Bensby_2014};
we however caution this comparison given that the Bensby sample does not extend to the most metal-poor tail of Nyx. The differential abundances of the metal-poor Nyx stars generally have larger uncertainties ($\sigma \sim 0.2$) compared to the metal-rich ones. These uncertainties are dominated by systematic effects, since the differential abundances of metal-poor stars are more susceptible to stellar parameter (mainly $\teff$ and $\log g$) uncertainties. The Na abundances of the metal-poor Nyx stars are slightly higher than the most metal-poor stars of the Bensby sample, on average by 0.2~dex. The $\alpha$-element abundances of the metal-poor Nyx stars are within 0.2~dex of the thick disk.

Fig.~\ref{chemical_abundances1} demonstrates that one of the metal-poor stars (Nyx175906) has a relatively high ratio of $\mbox{[Ba/Y]}$. Unlike the high $\mbox{[Ba/Y]}$ characteristic of stars in dwarf galaxies, this stems from high $\mbox{[Ba/Fe]}$ abundances, rather than low $\mbox{[Y/Fe]}$ abundances, and is not considered indicative of a low mass dwarf galaxy origin.

\section{Kinematics of Nyx Stars}
\label{sec:dynamics}

\begin{figure*}[t]
\begin{center}
\includegraphics[width=0.95\textwidth]{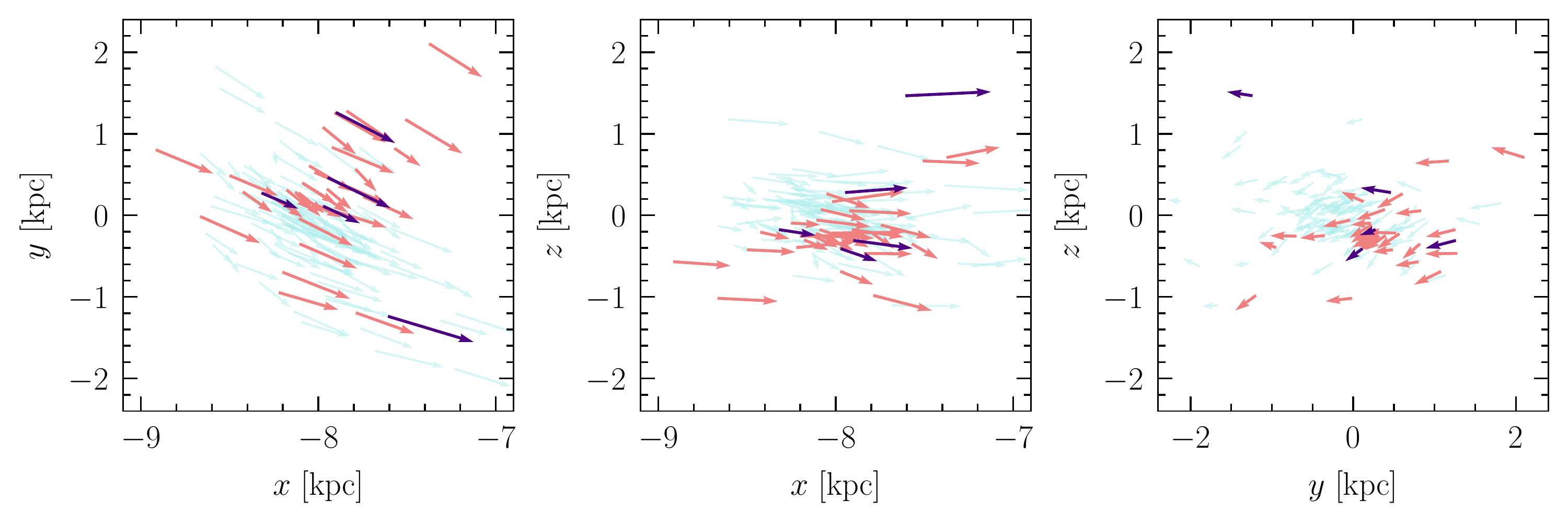} \\
\end{center}
\caption{
Spatial distributions of Nyx stars observed with high-resolution spectroscopy in this paper. The velocities are shown as arrows, where the length of the arrow is proportional to the speed of the stars in that projection (e.g., in the first panel, the length of the arrow is proportional to $\sqrt{v_x^2 + v_y^2}$). The observed stars in this paper are shown in peach, with the metal-poor stars in purple. Overlaid in light turquoise are the rest of the Nyx stars, selected as having probabilities $>0.75$ of belonging to the Nyx velocity Gaussian distribution \citep[][see Sec.~\ref{sec:obs} for more details on the selection of the stars]{2020NatAs...4.1078N}. These stars have been subsampled by a factor of five for clarity of the figure. The metal-poor stars (purple) are coherent with the rest of the sample.}
\label{fig:spatial_distributions}
\end{figure*}

The Nyx stream has peculiar kinematics for disk stars, with a slow prograde velocity of $\langle v_\phi \rangle = 130$ \kms, compared to the thick disk at $\langle v_\phi \rangle \sim 200$ \kms \citep{2018A&A...616A..11G}, and a large radial velocity component of $\langle v_r \rangle = 134$ \kms \citep{2020NatAs...4.1078N}, compared to the thick disk at $\langle v_r \rangle \sim$ 1\kms \citep{2021A&A...649A...1G,Vieira_2022}. 
Fig.~\ref{chemical_abundances1} shows that while the majority of the Nyx stars have similar chemical abundances to the thick disk, five stars populate a metal-poor tail that could be part of a different component. We now re-examine the Nyx star kinematics to see if these metal-poor stars are also kinematically distinct.

Our kinematic quantities for Nyx (including the unobserved sample) and the \citet{Bensby_2014} thick disk stars are determined from \Gaia eDR3 \citep{2021A&A...649A...1G}\footnote{We have checked that the absolute speed distribution ($\Vert v_r, v_{\phi}, v_{\theta}\Vert$) from \Gaia~DR3 \citep{Gaia_DR3} is within $5\%$ of that from \Gaia~eDR3} with radial velocities from \Gaia DR2 \citep{Gaia_DR2}. Velocities in Cartesian coordinates are based on the catalog by \citet{2021MNRAS.503.1374M}.\footnote{\url{https://sites.google.com/view/tmarchetti/research}}
We use \texttt{Galpy}\footnote{\url{http://github.com/jobovy/galpy}} \citep{2015ApJS..216...29B} to derive the eccentricity $e$, maximum vertical distance $z_{\rm{max}}$, and apocenter of the stellar orbits, assuming the \texttt{MWPotential2014} potential and integrating each orbit backwards 1 Gyr in steps of 1 Myr.\footnote{We adopt a right-handed coordinate system in this analysis.}
Fig.~\ref{fig:spatial_distributions} shows the positions of Nyx stars studied in this paper in Galactocentric Cartesian coordinates; the metal-rich stars are indicated in peach, and the five metal-poor stars are highlighted in purple. The velocities of the stars are shown as arrows. The blue background arrows show the kinematics of the rest of the Nyx stars identified by \citet{2020NatAs...4.1078N} with neural network scores $S>0.95$ (i.e. confidence that the neural network has in labeling the star as accreted) and high ($>0.75$) probabilities of belonging to the Nyx best-fit velocity distribution.
We find that the direction of the observed metal-poor stars are mostly coherent with the rest of unobserved Nyx stars, where all the stars display the signature prograde behavior of Nyx kinematics. 
There is no obvious distinction based on metallicity on the spatial distribution of the stars.

\begin{figure*}[t]
\begin{center}
\includegraphics[width=0.95\textwidth]{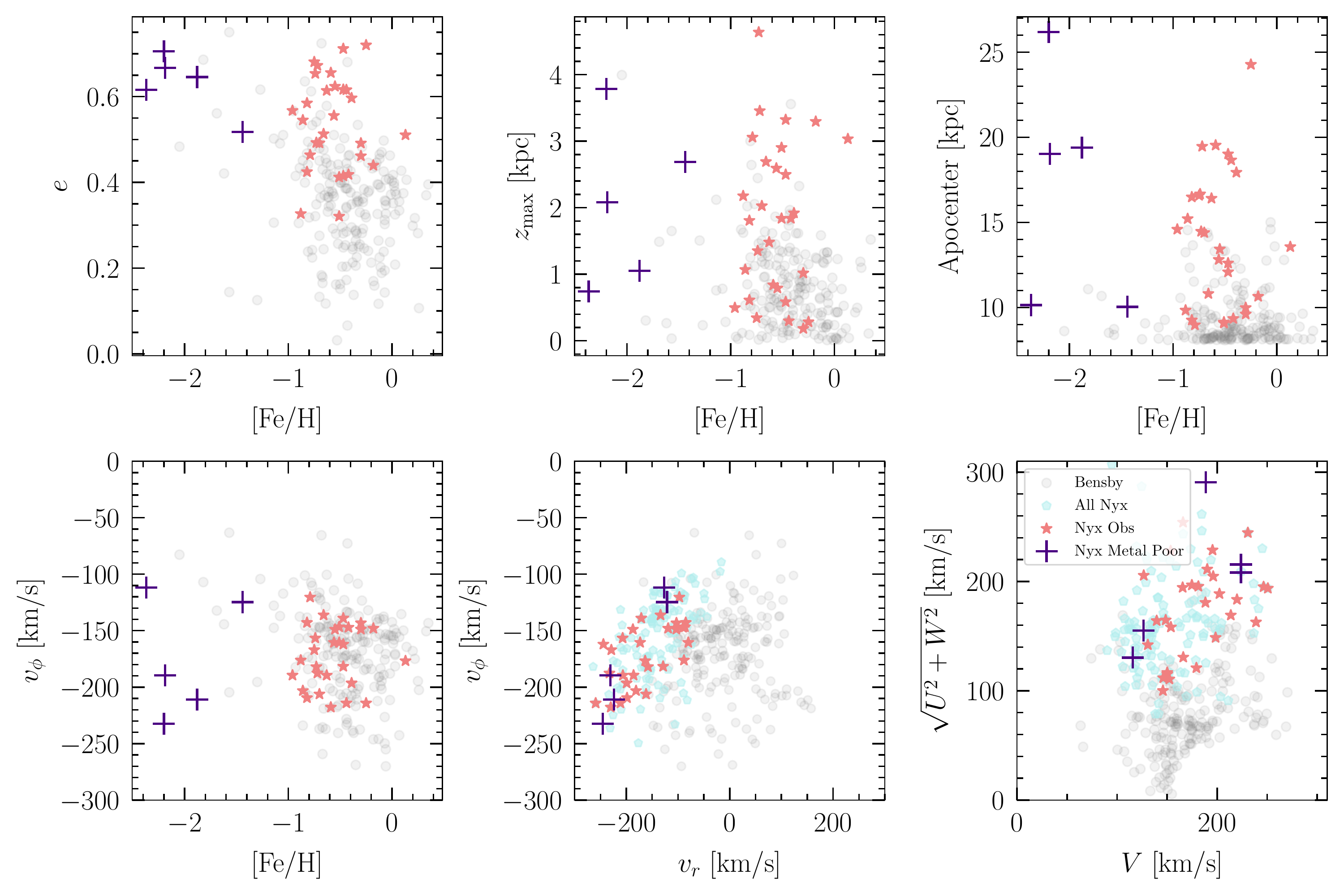} \\
\end{center}
\caption{Orbital parameters including eccentricities $e$, maximum vertical distances from the Galactic plane $z_\mathrm{max}$, apocenters, and $v_\phi$ versus metallicities $\mbox{[Fe/H]}$ of the Nyx stars, as well as a kinematics plot $v_r - v_\phi$ in Galactocentric spherical coordinates, and a Toomre plot of the Nyx stars. The Bensby thick disk stars are denoted by the light gray circles in the background. The rest of the unobserved sample of Nyx stars is shown in cyan (see Fig.~\ref{fig:spatial_distributions} for details on the selection of the Nyx stars). These stars are not shown in the orbital parameter panels as we do not have matching metallicity measurements.}
\label{fig:orbital_parameters}
\end{figure*}

Fig.~\ref{fig:orbital_parameters} shows the derived orbital parameters of the Nyx and thick disk stars, with the metal-poor stars highlighted in purple. The top panels of Fig.~\ref{fig:orbital_parameters} show the derived orbital parameters from \code{Galpy}. The bottom left panel shows the rotational velocity $v_\phi$ as a function of the metallicity. The bottom middle panel shows the Galactocentric spherical velocities of the observed Nyx stars (peach stars), the metal-poor stars (purple crosses), and the rest of the Nyx stars (cyan pentagons) compared to the reference stars from \citet{Bensby_2014} (gray circles) in $v_r-v_\phi$ space, while the bottom right panel shows the Toomre diagram of the same stars. Note that the values of $v_r$ and $v_\phi$ have the opposite sign from \citet{2020NatAs...4.1078N}.

As shown in Fig.~\ref{fig:orbital_parameters}, we find systematic differences between the orbital parameters of the Nyx stars and the thick disk stars. On average\footnote{We avoid a detailed quantitative comparison with the Bensby sample as the latter might not reflect the full distribution of the thick disk. We therefore only quote the mean values of the Bensby distributions.} the eccentricities of the thick disk stars are lower than those of the Nyx stars; the average thick disk eccentricity is $\sim 0.4$, while that of all the observed Nyx stars\footnote{The numbers quoted are the median and the 16th and 84th percentiles.} is $0.56^{+0.10}_{-0.13}$, and that of only the five metal-poor Nyx stars is $0.65^{+0.04}_{-0.06}$.
Note that \citet{Bensby_2014} have a complex selection function, but their eccentricity distribution is similar to those found in more representative surveys \citep[e.g.,][]{Yan_2019}.
The maximum vertical distance of these orbits is on average $0.7$~kpc for the thick disk stars, while it is higher for the Nyx stars: $1.84^{+1.21}_{-1.25}$~kpc for the entire Nyx sample and $2.08^{+1.00}_{-1.14}$ kpc for the metal-poor subset.
In the top right panel of Fig.~\ref{fig:orbital_parameters}, we find that the apocenters of the thick disk stars are smaller than those of Nyx stars: $\sim 9$ kpc for the thick disk, compared to the larger values of $14.0^{+5.0}_{-4.3}$ kpc and $19.1^{+2.8}_{-8.9}$ kpc for the observed Nyx sample and the metal-poor subset, respectively.

In the bottom left panel of Fig.~\ref{fig:orbital_parameters}, we show the distribution of rotational velocity as a function of metallicity of the Nyx and Bensby stars. We find that the metal-rich Nyx stars are consistent with the Bensby stars. However, the five metal poor stars split into three stars with high rotational velocity of $\sim 200$~km/s, and two stars with much smaller rotational velocity $\sim 120$~km/s, which will be further discussed in Sec.~\ref{sec:discussion}.

In the bottom middle panel of Fig.~\ref{fig:orbital_parameters}, we find that the five metal-poor stars are not clustered in exactly the same region of velocity space, but are rather on the edges of the distribution, where three of the stars are on the most radial orbits, while the other two stars are two of the slowest rotating stars out of the observed Nyx sample. 
The overall kinematics of the Nyx stars, shown in the Toomre plot as the right bottom panel of Fig.~\ref{fig:orbital_parameters}, are mostly inconsistent with those of thick disk stars, especially the metal-poor tail. The thick disk stars tend to have lower $\sqrt{U^2 + W^2}$ values than the Nyx stars, which is consistent with the high radial velocity of the Nyx stars. 

To address the discrepancies between the Bensby and Nyx samples, one has to take into account the selection effects of both samples. First, we estimate low contamination rates in the Nyx sample formed \textit{in-situ} based on the neural network classification \citep{2020A&A...636A..75O}, which are expected to be close to zero. However, this estimate could be artificially low due to bias in the simulations used to train the neural network, and is rather a reflection of the simulation's galactic formation. Second, the Bensby sample has a complex selection function, and a $3.0\%$ halo contamination rate; \citet{Bensby_2014} provided probability ratios that reflect the likelihood of belonging to the halo compared to the thick disk. Such probabilities are based purely on kinematics and therefore assume distributions of thick disk, which could include some structure like Nyx, albeit probably as a small fraction. It is therefore important to keep these systematic differences in mind during this comparative study.

\section{Discussion: On the Origin of Nyx}
\label{sec:discussion}

The main result of this paper is that most Nyx stars are metal-rich ($\mbox{[Fe/H]} > -1$) and have detailed chemical compositions similar to the thick disk.
The chemical similarity to the thick disk, and in particular the lack of chemical signatures indicating relative inefficient star formation expected in dwarf galaxies (low-$\alpha$ and high $\mbox{[Ba/Y]}$ or $\mbox{[La/Y]}$; \citealt{Venn_2004}), clearly shows that Nyx is not composed of one relatively massive and metal-rich dwarf galaxy.
A high-$\alpha$ dwarf galaxy would have had to accrete very early into the Milky Way. The two most significant early mergers into the Milky Way are \Gaia Sausage Enceladus \citep{Belokurov_2018,Helmi_2018} and the inner galaxy structure Kraken/Heracles \citep{Kruijssen_2020,Horta_2020}. 
These systems have $\mbox{[Fe/H]} < -1$, while Nyx has an average metallicity $\mbox{[Fe/H]}\sim -0.5$, and thus if it was a dwarf galaxy it would be substantially more massive than GSE and Kraken/Heracles.
We thus disfavor the scenario where Nyx is a single massive accretion, agreeing with recent work by other authors using chemical abundances from large spectroscopic surveys \citep{Zucker_2021, Horta_2022}, though we have added the extra information from neutron-capture elements.
Nyx thus might not be able to make a significant contribution to the dark matter local phase space distribution: either this stream is due to some baryonic process, or it is from mergers that are too low mass to deposit a large amount of dark matter \citep[e.g.,][]{Read_2008, Pillepich_2014}; although the latter scenario requires further investigation since it depends on properties of the mergers, including for example their orbital parameters.  

However, it still remains unclear how to produce a stream like Nyx, especially given that it has a significant metal-poor component and spans $-2.5 \lesssim \mbox{[Fe/H]} \lesssim -0.5$.
We thus briefly consider other possible origins for Nyx. The key observations to explain are: (1) the prograde orbit within the thick disk volume, (2) the relatively high eccentricity of Nyx stars ($0.4 < e < 0.8$), (3) the chemical similarity between Nyx and the thick disk, and (4) the presence of a metal-poor Nyx population.
We note that many of the scenarios discussed below are similar to formation scenarios of the thick disk (whose origin is also still debated), but we consider them here in the context of needing to explain a high eccentricity subset of the disk.

\subsection{Major Mergers}
We have shown that the stars in Nyx itself cannot be from a large accretion event, but a large accretion event could still have impacted an existing disk in such a way as to produce the eccentric Nyx stars.
There are two main ways this could occur.
First, a massive merger could dynamically perturb disk stars from circular orbits to hotter orbits and higher eccentricities \citep[e.g.,][]{Villalobos_2008}. The subdominant metal-poor stars originating from the dwarf galaxy could explain the metal-poor Nyx component, and the thick disk would have a range of eccentricities at similar composition that could explain the metal-rich Nyx component. \citet{Sales_2009} pointed out that the average eccentricity of the perturbed thick disk stars would only be $e \sim 0.2$, which disfavors this explanation for Nyx, though additional simulations could be useful.
Chemically, it appears that Nyx could be a part of the in-situ halo \citep[e.g.,][]{Bonaca_2017, Haywood_2018}, a high-eccentricity extension of the high-$\alpha$ disk population which possibly originated from a dynamically-heated disk \citep[e.g.,][]{Naidu_2020}. The metal-poor Nyx stars could be a part of the metal-poor tail of the distribution. The broad, continuous eccentricity distribution of the in-situ halo could explain the high eccentricity of Nyx stars.
However, we do not consider the evidence conclusive, since it remains unclear as to why there should be an additional overdensity in the kinematic space for Nyx overlapping with where the in-situ halo normally lies. 

Alternatively, a massive gas-rich merger could be responsible for Nyx.
It has been proposed that the thick disk could be formed during a turbulent epoch of gas-rich mergers, prior to the formation of the thin disk \citep{Brook_2004}. One reason this is an attractive explanation is that it could explain both the metal-rich and metal-poor stars in Nyx: the predominant metal-rich Nyx stars would have formed in the Milky Way using gas stripped from accreted satellites, while the subdominant metal-poor Nyx stars would come from the accreted galaxy.\footnote{Also see recent work \citep[e.g.][]{Santistevan_2021, Sestito_2021}}. However, because gas is collisional, it would end up in orbits distinct from the accreted stars. \citet{Sales_2009} found in simulations that the stars formed from the stripped gas would only have $e \sim 0.2$, too low to explain the behavior of the metal-rich Nyx stars. 
Therefore, we find it unlikely that the Nyx stars originated from a major merger event, be it through the dynamical heating scenario or the gas-rich merger scenario.

\subsection{Minor Mergers}
\citet{Abadi_2003} proposed that over 60\% of the Galactic thick disk could be stars accreted from multiple tidally disrupted satellite galaxies on prograde orbits (also found in more recent simulations, see \citealt{Mardini_2022}).
\citet{Sales_2009} showed that this scenario would result in relatively high eccentricities, $e \sim 0.5$, consistent with most observed Nyx stars.
A minor merger could provide the metal-poor Nyx stars, while producing the metal-rich Nyx stars by pulling a small fraction of disk stars into the same orbit.

The high-$\alpha$ abundances of the metal-poor Nyx stars challenge this idea: most lower-mass dwarf galaxies have low-$\alpha$ for their most metal-rich ($\mbox{[Fe/H]}\sim -1.0$) stars \citep{Venn_2004}, because maintaining high-$\alpha$ requires a high star formation efficiency that keeps the core-collapse to Type Ia supernova ratio high.
However, high-$\alpha$ can be maintained if the minor merger occurred relatively early.
Indeed, two candidate accretion events have high-$\alpha$: the inner galaxy structure Kraken/Koala/Heracles \citep{Kruijssen_2020, Forbes_2020,Horta_2020}, and the Atari Disk \citep[or metal-weak thick disk; e.g.,][]{Norris85,Chiba00,Carollo_2019,An_2020,Mardini_2022}.
Neither of these known events is the same as Nyx: Kraken is restricted to Galactocentric radii $r < 5$\,kpc while Atari has much lower mean radial velocity ($\langle v_r \rangle \sim 10\,\kms$).
We thus consider an early minor merger to be a plausible explanation for Nyx.
Note that a lower-mass merger could still result in a dark matter substructure, but would likely make a fairly negligible ($<10\%$) contribution to the local phase space distribution \citep{Pillepich_2014}.

\subsection{Radial Migration}

Resonant scattering of spiral arms may cause disk stars on near-circular orbits to move radially, while preserving their eccentricities \citep[e.g.,][]{Sellwood_2002, Roskar_2008}. Based on radial migration, \citet{Schonroch_2009} constructed chemical evolution models of the Milky Way that demonstrate the coevolution of the Galactic thick and thin disk. They found that the resulting $\mbox{\alphafe}$ versus $\mbox{[Fe/H]}$ trend closely resembled that from \citet{Venn_2004}, and displayed a dichotomy between the thick and thin disks. \citet{Ruchti2011} investigated the chemodynamics of the high-$\alpha$ metal-poor thick disk stars ($-2.5\lesssim \mbox{[Fe/H]} \lesssim -0.7$) from RAVE \citep{Steinmetz_2006}, and considered radial migration as one of the possible origins of those stars. Thus, radial migration might be able to produce a predominant metal-rich stellar population chemically similar to the thick disk and a subdominant high-$\alpha$ metal-poor stellar population.
However, radial migration generally results only in moderate eccentricities of $e \sim 0.2$ (as well as an eccentricity cut-off at $e \sim 0.6$, \citealt{Sales_2009}), while those of the Nyx stars are $0.56^{+0.10}_{-0.13}$.

\subsection{Relic of Thick Disk Formation}

Several recent studies have attempted to study the metal-poor \emph{in-situ} component of the Milky Way with spectroscopic surveys \citep{Xiang_2022, Belokurov_2022, Conroy_2022}. These studies have broadly concluded that between $-2.5 \lesssim \mbox{[Fe/H]} \lesssim -0.5$ (corresponding to an age ${\sim}13$ Gyr ago), the Milky Way both rapidly spun up (from $\langle v_\phi \rangle < 100\kms$ up to $\langle v_\phi\rangle \sim 175\kms$) and greatly increased its star formation efficiency before merging with GSE. In this early turbulent gas-rich disk, it is plausible that a subset of \emph{in-situ} disk stars could form with high eccentricities \citep[e.g.,][]{van_Donkelaar_2022}.

Nyx could potentially be a remnant of this early time, either a distinct phase of formation or a high-eccentricity tail of the overall thick disk formation.
Indeed, the Nyx stars span the range of rotational velocities (bottom left panel of Fig.~\ref{fig:orbital_parameters}) and metallicities expected for this scenario.
A challenge for this scenario is that on average, the azimuthal velocities are expected to increase with metallicity \citep[see e.g.]{Belokurov_2022}, while in Nyx many of the fastest rotating stars are the most metal-poor stars (Figure~\ref{fig:orbital_parameters}).
Still, given uncertainties in the details of early Milky Way formation, we consider this a promising explanation.
An obvious conclusion of this scenario is that structures like Nyx should be present at any position within the disk, not just the Solar Neighborhood.

\section{Conclusions}
\label{sec:conclusion}
We analyzed the chemical abundances and kinematics of 34 Nyx stars observed with Magellan/MIKE or Keck/HIRES. These high purity stars were selected from 94 Nyx stars in the accreted star catalog defined by \citet{2020NatAs...4.1078N}. In order to determine the origin of the Nyx stars, we provided detailed chemodynamic comparison of these Nyx stars to thick disk stars from \citet{Bensby_2014}.
We employed a differential abundance analysis to study the chemical compositions of the Nyx stars. 

Our analysis reveals that the chemical abundances of the Nyx stars are mostly consistent with the high-$\alpha$ thick disk stars, with the exception of five stars in the metal-poor tail ($\mbox{[Fe/H]} \sim -2.0$). 
The abundance results we attain are of much higher precision than those from RAVE-on \citep{Casey_2017}, used in the Nyx discovery paper \citep{2020NatAs...4.1078N}, and show that the metal-rich Nyx stars are chemically consistent with thick disk stars. Both metal-poor and metal-rich Nyx stars have overall similar kinematic properties, although the metal-poor Nyx stars have somewhat higher eccentricities ($0.5 < e < 0.7$) than the metal-rich Nyx stars ($e \sim 0.5$). 

The chemical abundances clearly rule out Nyx as the result of a single massive dwarf galaxy merger.
However, it is still not known how to produce a population of stars with identical chemistry to the thick disk but such high eccentricities.
We consider multiple formation scenarios, finding the most likely scenarios are that Nyx is the signature of an early minor merger or that Nyx is an unusual phase in the Milky Way's thick disk formation history. 
The former scenario---Nyx originated from multiple small dwarf galaxy mergers---would require that the merger event occurred relatively early,
and would likely only make a fairly negligible ($<10\%$) contribution to the dark matter local phase space distribution; whereas the latter scenario---Nyx is the relic of the early thick disk formation history---might be challenged by the fact that the fastest rotating Nyx stars are the most metal-poor ones. The latter scenario clearly suggests that similar structures like Nyx exist should exist elsewhere within the disk.

Future simulations and observations might shed light on the possible formation scenario of the Nyx stream. Specifically, isolated simulations of early accretion events between the Milky Way's primordial disk and multiple satellite dwarf galaxies could explore the formation history of the Galactic thick disk. Recent advances, particularly in simulations of stable disk formation and disk perturbations, open the path to future studies of Nyx-like structures and their origins \citep[e.g.,][]{Timmy_2022}. A detailed chemodynamic study of the stars produced in such streams would help determine whether the metal-rich Nyx stars and the high eccentricity metal-poor tail formed through multiple minor dwarf galaxy mergers. If simulations can produce stars with high-$\alpha$ abundances from early mergers, it could provide evidence for this scenario.

Alternatively, if Nyx is a remnant of the early turbulent Milky Way formation, future \Gaia data releases and spectroscopic surveys such as SDSS-V, WEAVE, and 4MOST \citep{Kollmeier2017,Dalton2012,deJong2016} should identify similar structures further away from the Solar Neighborhood.
A combination of high-resolution simulations with current and future observations should thus aid in piecing together the origin of Nyx and understanding its role in the formation and structure of our Galaxy.

\begin{acknowledgments}

We are grateful to E. Holmbeck and R. Naidu for their contributions to the observation of Nyx stars. We thank the referee for comments that significantly improved this paper. SW thanks R. Naidu, P. Re Fiorentin, and A. Wetzel for helpful discussions. LN and ML thank T. Cohen and B. Ostdeik for their early collaboration and the Nyx discovery. 

APJ acknowledges support from NSF grant AST 2206264.
ML is supported by the DOE under Award Number DE-SC0007968. MACdlR is supported by a Stanford Science Fellowship.

This work presents results from the European Space Agency (ESA) space mission \Gaia. \Gaia data are being processed by the \Gaia Data Processing and Analysis Consortium (DPAC). Funding for the DPAC is provided by national institutions, in particular the institutions participating in the \Gaia MultiLateral Agreement (MLA). The Gaia mission website is \url{https://www.cosmos.esa.int/gaia}. The Gaia archive website is \url{https://archives.esac.esa.int/gaia}. This research makes use of public auxiliary data provided by ESA/Gaia/DPAC/CU5 and prepared by Carine Babusiaux.

This research has made use of the NASA/IPAC Infrared Science Archive, which is funded by the National Aeronautics and Space Administration and operated by the California Institute of Technology.

This research has made use of the Keck Observatory Archive (KOA), which is operated by the W. M. Keck Observatory and the NASA Exoplanet Science Institute (NExScI), under contract with the National Aeronautics and Space Administration.
This paper includes data gathered with the 6.5 meter Magellan Telescopes located at Las Campanas Observatory, Chile.

\end{acknowledgments}

\facilities{Keck (HIRES), Magellan (MIKE), \Gaia}

\software{
\code{CarPy} \citep{Kelson_2003},
\code{smhr} \citep{Casey_2014,Ji_2020},
\code{numpy} \citep{numpy}, 
\code{scipy} \citep{scipy},
\code{matplotlib} \citep{matplotlib}, 
\code{astropy} \citep{astropy:2013,astropy:2018},
\code{AGAMA} \citep{vasiliev19}
          }

\bibliography{main}{}

\bibliographystyle{aasjournal}

\appendix

\section{Absolute Abundances and Uncertainties}
\label{sec:abs}
We determine the absolute chemical abundances of the observed Nyx stars---i.e., assuming that we have accurate atomic data, model atmospheres, and radiative transfer physics.
The uncertainties of the absolute abundances are determined following \citet{Ji_2020}, i.e., propagating spectrum noise and stellar parameter uncertainties into the final results with a weighted average, where the weights generalize the usual inverse-variance weight under the assumption that individual line abundances are correlated due to stellar parameters.
There exist significant zero-point offsets ($\delta \sim 0.2$) between the absolute abundances of the Nyx stars and literature \citep[see e.g.,][]{Bensby_2014}.

Here, we discuss the absolute abundances of Al~I, K~I, Sc~II, V~I, Mn~I, and Eu~II. Al and K abundances are determined using equivalent width measurement, while Sc~II, V~I, Mn~I and Eu~II abundances are determined using spectral synthesis. These abundances are not determined differentially for the following reasons: all Al lines and the Eu line at 6645 fall into chip gaps for the Nyx stars observed with Keck/HIRES, and are only measured in the six stars observed with Magellan/MIKE;\footnote{Although bluer Eu lines are visible at 4129~\AA~and 4205~\AA, these wavelength regions are too blended for accurate abundance analysis.} while K, Sc, V and Mn abundances were not determined by \citet{Bensby_2014}.

\begin{figure*}
\begin{center}
\includegraphics[width=\textwidth]{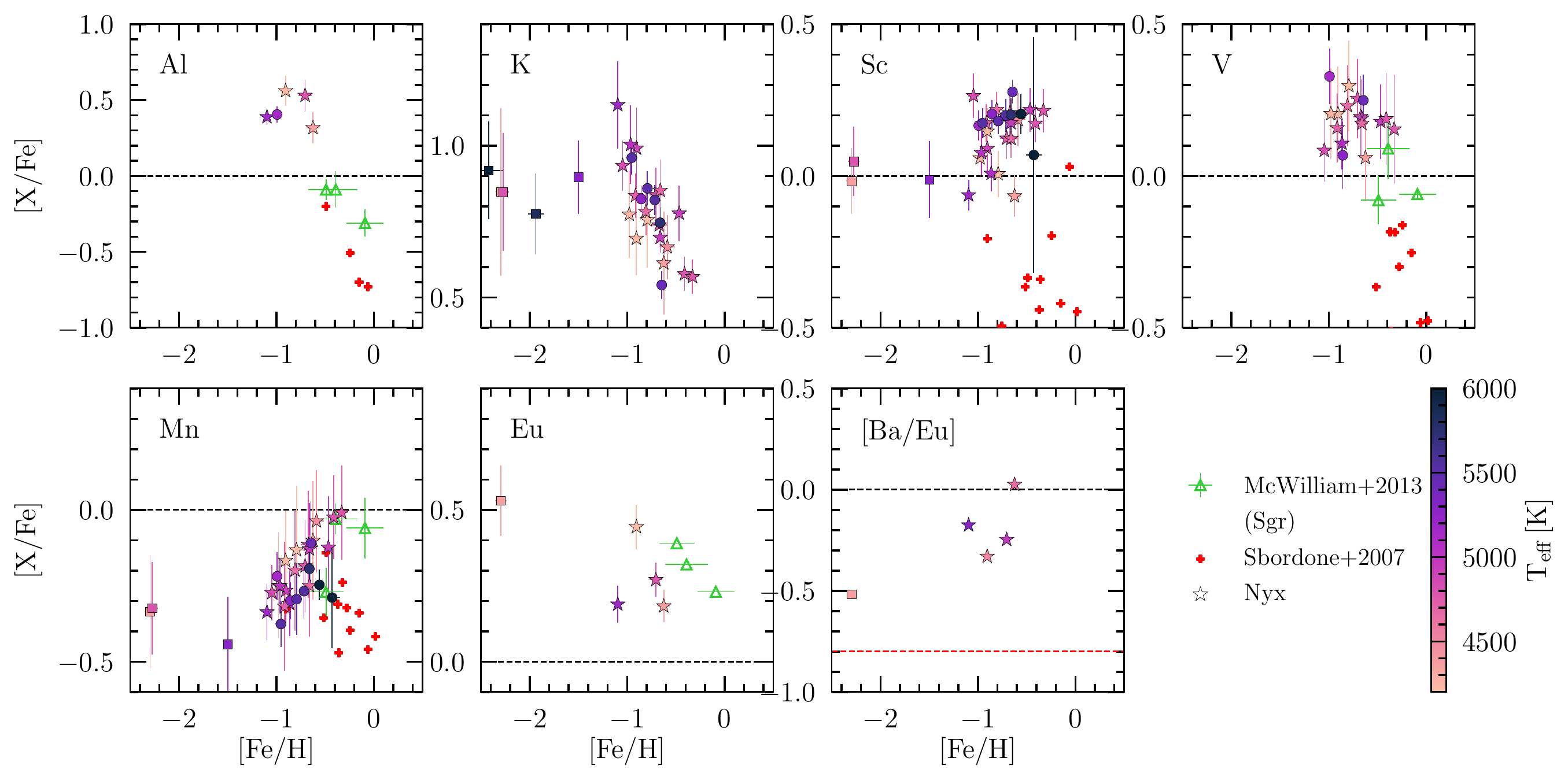} \\
\end{center}
\caption{Absolute chemical abundances ($\mbox{[X/Fe]}$ versus $\mbox{[Fe/H]}$, y-axis is $\mbox{[X/Fe]}$ except the panel where another ratio is indicated.) of the observed Nyx stars, color-coded by effective temperature, $\teff$. The effective temperatures range from $4200$ to $6000\,\unit{K}$. For comparison, three Sgr dwarf galaxy stars \citet{McWilliam_2013} are shown by green triangles. 12 Sgr dwarf galaxy giants from \citet{Sbordone_2007} are indicated by red plusses. A dashed line at $\mbox{[X/Fe]} = 0.0$ is shown in each panel as a reference. A dashed red line at $\mbox{[Ba/Eu]} = -0.8$ indicates the pure r-process ratio \citep{Sneden_2008}. In general, the chemical abundances of the metal-rich Nyx component are distinct from the Sgr stars. }
\label{chemical_abundances4}
\end{figure*}

Fig.~\ref{chemical_abundances4} shows the absolute abundances of the Nyx stars for these six elements, color-coded by effective temperatures. 
The abundances of three Sagittarius~(Sgr) dwarf galaxy stars from \citet{McWilliam_2013} and 12 Sgr giant stars from \citet{Sbordone_2007} are also shown for comparison.

As shown in Fig.~\ref{chemical_abundances4}, 
The odd-Z (Al, Sc, V) abundances of Sgr stars are significantly lower (by $\sim0.6$~dex) than the metal-rich Nyx stars, although their metallicities are slightly higher.
The iron-peak (Mn) abundances have relatively small dispersion ($\sim0.1$~dex). $\mbox{[Mn/Fe]}$ displays a slightly upward trend with respect to $\mbox{[Fe/H]}$. 
Fig.~\ref{chemical_abundances4} also shows the $\mbox{[Ba/Eu]}$ ratios of the Nyx stars, an indicator of the relative amount of s-process and r-process enrichment. The $\mbox{[Ba/Eu]}$ ratios increase with metallicity from a pure r-process ratio of $\mbox{[Ba/Eu]}=-0.8$ \citep{Sneden_2008} to a fairly s-process dominated ratio $\mbox{[Ba/Eu]}\gtrsim 0.0$, following the rise of the s-process \citep{Simmerer_2004}.
One of the five Nyx stars in the metal-poor tail (Nyx122147) has a detectable Eu abundance of $\mbox{[Ba/Fe]}=0.572$ and $\mbox{[Ba/Eu]}=-0.515$, indicating r-process domination with some s-process influence.
In Fig.~\ref{chemical_abundances4}, the overall $\mbox{[Eu/Fe]}$ versus $\mbox{[Fe/H]}$ trend is flat, matching the Milky Way at these metallicities \citep{Battistini_2016} ($\mbox{[Eu/Fe]}\sim 0.3$). There are too few Eu measurements to robustly measure the expected r-process scatter.

\end{document}